\documentclass{IEEEtrans}

\usepackage{cite}
\usepackage{amsmath,amssymb,amsfonts}
\usepackage{graphicx,color}
\usepackage{textcomp}

\usepackage{xcolor}
\usepackage{verbatim}
\usepackage{multirow}
\usepackage{tabularx}
\usepackage{hhline}
\usepackage{algorithm}
\usepackage{algpseudocode}
\usepackage{footnote}
\usepackage[normalem]{ulem}
\usepackage{orcidlink}
\makesavenoteenv{tabular}
\makesavenoteenv{table}

\newcolumntype{Y}{>{\centering\arraybackslash}X}

\AtBeginDocument{\definecolor{ojcolor}{cmyk}{0.93,0.59,0.15,0.02}}

\def\BibTeX{{\rm B\kern-.05em{\sc i\kern-.025em b}\kern-.08em
    T\kern-.1667em\lower.7ex\hbox{E}\kern-.125emX}}

\begin{document}

\title{Towards a Multi-Layer Defence Framework for Securing Near-Real-Time Operations in Open RAN}

\author{
    \IEEEauthorblockN{
        Hamed~Alimohammadi\orcidlink{0000-0001-6005-691X}\IEEEauthorrefmark{1},
        Samara~Mayhoub\orcidlink{0000-0001-7629-0532}\IEEEauthorrefmark{2},
        Sotiris~Chatzimiltis\orcidlink{0000-0002-4980-2608}\IEEEauthorrefmark{1},~\IEEEmembership{Graduate Student Member,~IEEE},\\
        Mohammad~Shojafar\orcidlink{0000-0003-3284-5086}\IEEEauthorrefmark{1},~\IEEEmembership{Senior Member,~IEEE},
        and~Muhammad~Nasir~Mumtaz~Bhutta\orcidlink{0000-0001-5580-4282}\IEEEauthorrefmark{3},~\IEEEmembership{Member,~IEEE}
    }\\
    \IEEEauthorblockA{
        \IEEEauthorrefmark{1}6G Innovation Centre (6GIC), Institute for Communication Systems, University of Surrey, Guildford, UK \\
        Email: \{h.alimohammadi, sc02449, m.shojafar\}@surrey.ac.uk\\
    }
    \IEEEauthorblockA{
        \IEEEauthorrefmark{2}Centre for Secure Information Technologies (CSIT), Queen’s University Belfast, Belfast, UK \\
        Email: s.mayhoub@qub.ac.uk\\
    }
    \IEEEauthorblockA{
        \IEEEauthorrefmark{3}Computer Science and Information Technology Department, College of Engineering,\\
        Abu Dhabi University, United Arab Emirates \\
        Email: muhammad.bhutta@adu.ac.ae
    }
}

\markboth{Alimohammadi \textit{et al.}: Towards a Multi-Layer Defence Framework for Securing Near-Real-Time Operations in Open RAN}{}

\maketitle


\begin{abstract}
Securing the near-real-time (near-RT) control operations in Open Radio Access Networks (Open RAN) is increasingly critical, yet remains insufficiently addressed, as new runtime threats target the control loop while the system is operational. In this paper, we propose a multi-layer defence framework designed to enhance the security of near-RT RAN Intelligent Controller (RIC) operations. We classify operational-time threats into three categories—message-level, data-level, and control logic-level—and design and implement a dedicated detection and mitigation component for each: a signature-based E2 message inspection module performing structural and semantic validation of signalling exchanges, a telemetry poisoning detector based on temporal anomaly scoring using an LSTM network, and a runtime xApp attestation mechanism based on an execution-time hash challenge–response. The framework is evaluated on an Open RAN testbed comprising FlexRIC and a commercial RAN emulator, demonstrating effective detection rates, low latency overheads, and practical integration feasibility. Results indicate that the proposed safeguards can operate within near-RT time constraints while significantly improving protection against runtime attacks, introducing less than 80 ms overhead for a network with 500 User Equipment (UEs). Overall, this work lays the foundation for deployable, layered, and policy-driven runtime security architectures for the near-RT RIC control loop in Open RAN, and provides an extensible framework into which future mitigation policies and threat-specific modules can be integrated.
\end{abstract}


\section{Introduction}
\label{sec:intro}

\IEEEPARstart{T}{he} transition to Open Radio Access Networks (Open RAN) represents a paradigm shift in how mobile networks are designed, deployed, and managed. By disaggregating traditional vendor-specific components and introducing open interfaces, Open RAN enables the integration of multivendor solutions, fosters innovation through third-party applications, and promotes programmability through intelligent control loops. A central pillar of this architecture is RAN Intelligent Controller (RIC), which is logically split into two parts: non-real-time (non-RT) RIC, operating at timescales above one second and typically located in the Service Management and Orchestration (SMO) layer, and near-real-time (near-RT) RIC, which functions at sub-second latencies (typically 10 ms to 1 s) to enable dynamic control of RAN behaviour. RIC executes control logic provided by third-party applications known as rApps and xApps, deployed on non-RT and near-RT RICs, respectively \cite{bib-1}.

In the Open RAN architecture, the most commonly adopted functional split is the 3GPP New Radio (NR) Split 7.2. This split disaggregates a base station into a Central Unit (CU), a Distributed Unit (DU), and a Radio Unit (RU). Furthermore, the CU is itself divided into a Control Plane (CU-CP) and a User Plane (CU-UP) component. In the Open RAN context, these are referred to as O-CU, O-DU, and O-RU to emphasise openness and interoperability.

The openness and disaggregation introduced by Open RAN increase flexibility and vendor diversity, but also expand the attack surface compared to traditional vertically integrated RAN architectures. Particularly, as near-RT RIC has an immediate impact on RAN operations, its security and trustworthiness have become critical concerns. The near-RT RIC communicates with E2 nodes, O-CUs and O-DUs, through the standardised E2 interface. xApps hosted on the near-RT RIC interact with these components to perform fine-grained, low-latency control and optimisation actions. Therefore, the near-RT RIC occupies a privileged position in the RAN control loop by receiving telemetry and signalling messages from E2 nodes, running inference or rule-based logic, mainly through its xApps, and issuing control actions that can reconfigure radio parameters, manage mobility, or optimise resource allocation. This dynamic and responsive behaviour, while offering substantial performance benefits, also creates new attack surfaces.

While current Open RAN security efforts mostly address earlier stages of the lifecycle, such as secure onboarding and authorisation of xApps via frameworks like XRF \cite{10228961,10816173}, and authentication of E2 interface connections through certificate-based mechanisms \cite{bib-10}, these measures do not address threats that can bypass \textit{proactive} safeguards and exploit vulnerabilities during operation, when the near-RT RIC is actively processing live inputs to generate decisions. Such runtime threats remain largely unaddressed in existing literature and, importantly, they emerge at different points in the near-RT control loop: malicious or manipulated E2 signalling impacts the message path, falsified or poisoned KPM reports corrupt inference inputs, and tampered or unauthorised xApps alter the control logic itself. Because these attack vectors target \textit{distinct functional layers} of the execution pipeline and exploit different weaknesses, no single defence mechanism can provide comprehensive protection. Message validation alone cannot detect telemetry manipulation, and anomaly detection alone cannot prevent compromised control logic from issuing unauthorised actions. Collectively, these gaps highlight the need for dedicated runtime \textit{reactive} safeguards that operate within the near-RT RIC control loop, and specifically motivate a \textit{multi-layer} defence approach in which complementary safeguards are placed at the message-level, data-level, and control logic-level to ensure the trustworthiness and resilience of the near-RT RIC decision-making process.

In this study, we approach the problem from the perspective of the near-RT RIC and focus on securing its \textit{runtime control loop}. We categorise \textit{externally influenced} runtime threats—those that do not stem from compromises within the near-RT RIC platform—into three dimensions: \textit{message-level} threats involving malicious or manipulated E2 signalling, \textit{data-level} threats where adversaries inject falsified telemetry such as spoofed KPM reports, and \textit{control logic-level} threats arising from tampered or malicious xApps. While other components may contribute to overall system trustworthiness, these three categories collectively capture the primary externally influenced surfaces that directly affect near-real-time control decisions. The internal platform of the near-RT RIC can be secured using mechanisms defined in O-RAN Alliance specifications, such as \cite{oran-srcs, oran-LCM, oran-Near-xApps}; therefore, we assume they are trusted and out of scope. Building on this categorisation, we propose a modular framework composed of deployable detection components paired with \textit{policy-driven mitigation} units that operate within the near-RT RIC during live execution. The framework provides a structured methodology for mapping safeguards to each threat category, enabling systematic deployment of reactive defences and supporting extensibility toward emerging runtime threats. To validate its practicality, we demonstrate the framework’s feasibility through the implementation of one detection and mitigation mechanism for each threat category on an Open RAN testbed. This work therefore offers a practical step toward layered runtime assurance in near-RT RIC operations, through a threat-guided and policy-centric defence design.

This paper makes the following key contributions:

\begin{itemize}
    \item \textit{Threat-Driven Categorisation and Defence Design:}
    We develop a systematic categorisation of runtime threats to the near-RT RIC, grouping them into message-level, data-level, and control logic-level classes. Building on this, we propose a generalised guideline for mapping detection and mitigation components to each threat category, providing operators with a structured methodology for placing runtime safeguards within the near-RT control loop.

    \item \textit{Modular Multi-Layer Runtime Safeguards:}
    We design and implement three complementary runtime defence components aligned with the above threat classes:
    \begin{itemize}
        \item A signature-based E2 message inspector for detecting malicious signalling patterns.
        \item An ML-based telemetry poisoning detector that identifies falsified KPM inputs.
        \item A runtime xApp attestation mechanism that verifies the integrity of executing xApps.
    \end{itemize}
    Each component is paired with a policy-driven mitigation unit, enabling flexible and operator-configurable runtime responses without modifying detection logic.

    \item \textit{System Integration and Experimental Validation:}
    We integrate all safeguards into a cohesive multi-layer framework for the near-RT RIC and evaluate its performance on an Open RAN testbed comprising FlexRIC and a commercial RAN emulator. Experimental results demonstrate high detection accuracy with low latency overhead, confirming the feasibility of deploying the proposed framework within near-RT time constraints.
\end{itemize}

The remainder of the paper is organised as follows. Section~\ref{sec:related} reviews related works in the area of near-RT RIC security. Section~\ref{sec:architecture} presents the proposed system architecture, including threat categorisation and deployment strategy. Section~\ref{sec:components} details the integrated runtime defence components, including the signature-based E2 message inspector, KPM poisoning detection module, and runtime xApp attestation mechanism. Section~\ref{sec:evaluation} describes the implementation details and experimental results for each module, as well as a use case deployment including all the modules. Section~\ref{sec:discussion} discusses limitations and future directions for enhancing the framework. Finally, Section~\ref{sec:conclusion} concludes the paper.

\section{Related Works}
\label{sec:related}

The security of Open RAN, especially the near-RT RIC and its open interfaces, has drawn increasing attention in recent years. Several works examine vulnerabilities and propose defences, yet most remain limited to static or single-layer protection, leaving runtime trust across operational layers insufficiently addressed.

\subsection{Fundamental Architectural and Interface-Level Security}

Fundamental studies \cite{bib-3,bib-7,bib-8,bib-9,bib-10,bib-11} analyse security threats in Open RAN stemming from its disaggregated architecture, open interfaces, and AI/ML-based control loops. They identify broad vulnerabilities such as protocol manipulation, data poisoning, and unauthorised access. Among them, \cite{bib-3} and \cite{bib-10} focus on the E2 interface, detailing Denial-of-Service (DoS) and Man-in-the-Middle (MitM) threats linked to weak authentication and encryption. \cite{bib-9} discusses adversarial input attacks capable of degrading AI/ML model performance in xApps and rApps, motivating subsequent research on adversarial robustness.

More specific efforts address the E2 interface directly. \cite{10433004} demonstrates practical E2-based attacks, including repeated subscription responses and spoofed RMR table exploits that disrupt RIC operations. \cite{10907911} extends this by modelling E2AP exchanges through FSMs (Finite State Machine) and fuzzing to ensure syntactic and state correctness. Both studies enhance understanding of protocol-level integrity. In contrast, our framework introduces runtime E2 message inspection without ASN.1 decoding, detecting malicious signatures directly from encoded traffic. These works are therefore complementary as \cite{10433004} and \cite{10907911} secure structural and state consistency, while our approach adds fast inline semantic validation under near-RT constraints.

\subsection{xApp-Based Runtime Safeguards}
\label{xapp-used}
Several works employ xApps as active security enforcers within the RIC, focusing on intrusion detection or runtime adaptation, but concentrated on the threats arising from UEs and threatening the network, not specifically near-RT-RIC. ZTRAN \cite{10495907} adopts a zero-trust architecture composed of authentication, IDS, and secure slicing xApps for adaptive mitigation, while \cite{10620781} develops IDS and Secure Slicing xApps to detect and isolate malicious UEs dynamically. Although effective for monitoring, these approaches treat xApps as security tools, not as assets whose own runtime integrity or control logic requires protection.

\subsection{Data-Level Robustness and Telemetry Integrity}

Several studies address the robustness and integrity of data exchanged within the near-RT RIC, encompassing both ML model inputs and telemetry flows. RADAR \cite{10971998} enhances DRL-based resource allocation xApps with adversarial input sanitisation and retraining to sustain QoS under attack. While this approach enhances model resilience under adversarial conditions, its scope remains limited. RADAR implements localized multi-layer defences within a single DRL-based xApp, improving inference stability but lacking cross-xApp. 

ZT-RIC \cite{10976081} introduces functional encryption to enable privacy-preserving telemetry sharing within the RIC, ensuring confidentiality while maintaining operational functionality. Complementary approaches such as \cite{chatzimiltis2025xaillm} leverage explainable AI and large language models to enhance semantic understanding and UEs behaviour anomaly interpretation through Key Performance Measurement (KPM) data monitoring. Likewise, \cite{bib-15} and \cite{bib-19} improve the robustness of ML-based xApps against adversarial or poisoned data using FGSM (Fast Gradient Sign Method) and PGD (Projected Gradient Descent)-based adversarial training, while \cite{bib-20} and \cite{bib-21} formalise and standardise KPM data structures and threat analysis to improve telemetry consistency.

Collectively, these studies reinforce data-plane resilience across privacy, robustness, and interpretability dimensions. However, they primarily rely on model retraining or offline validation and do not perform live semantic verification or integrity attestation of telemetry streams. Our framework addresses this gap by incorporating temporal anomaly detection and semantic validation of runtime KPM data as part of its multi-layer safeguarding mechanism.

\subsection{Platform-Level and Configuration Vulnerabilities}

Platform-level research has primarily focused on hardening, configuration auditing, and access control mechanisms within RIC implementations. \cite{10613612} investigates threats affecting open interfaces and proposes IPsec encryption and Autoencoder-based defences against adversarial ML, primarily during training and deployment. \cite{10.1145/3643833.3656118} performs penetration testing on µONOS RIC, uncovering authentication flaws, insecure REST APIs, and weak access control configurations that can expose critical components to unauthorised access. Additionally, \cite{bib-12} experimentally evaluates the impact of E2-induced disruptions, such as message delays and losses, on xApp performance, revealing runtime fragility but offering no mitigation.

These studies address platform and configuration-level security concerns, emphasising the importance of interface protection and software assurance. However, they are largely preventive or static, lacking the dynamic, runtime validation capabilities required to maintain operational trust within the near-RT RIC.

\subsection{Summary and Research Gap}

Across all these domains, existing research focuses on message syntax validation, privacy preservation, adversarial robustness, or static configuration. As summarised in Table~\ref{tab:related_comparison}, prior works address isolated aspects of RAN security but lack unified, in-operation assurance across message semantics, telemetry integrity, and xApp logic trustworthiness. Our work fills this gap by proposing a multi-layer runtime safeguarding framework that integrates signature-based E2 message inspection, telemetry anomaly detection, and xApp runtime attestation. This unified design ensures continuous, low-latency defence across the near-RT RIC control loop. Note that Table~\ref{tab:related_comparison} only includes comparable related works and some other references have been excluded due to their different scope, as described in Section~\ref{sec:related}.\ref{xapp-used}.

\begin{table*}[t]
\centering
\caption{Comparative coverage of related works.}
\label{tab:related_comparison}
\renewcommand{\arraystretch}{1.15}
\setlength{\tabcolsep}{3pt}
\begin{tabular}{|p{1.2cm}|p{6cm}|c|c|>{\centering\arraybackslash}p{1.6cm}|}
\hline
\textbf{Ref.} & \textbf{Focus of the Work} & \textbf{E2 Msg. Integrity} & \textbf{Data / Telemetry Integrity} & \textbf{Ctrl. Logic Integrity} \\
\hline
\cite{10433004}, \cite{10907911} & E2 interface integrity and protocol correctness—practical attacks (subscription spoofing, RMR exploits) and E2AP FSM/fuzzing for syntax and state validation. & \checkmark (Structural / Syntactic) & $\times$ & $\times$ \\
\hline

\cite{10971998} & RADAR: DRL-based xApp incorporating adversarial input sanitisation, retraining, and distillation for inference robustness. & $\times$ & \checkmark (Local model robustness) & $\times$ \\
\hline
\cite{10976081} & ZT-RIC: functional encryption for privacy-preserving telemetry within the near-RT RIC. & $\times$ & \checkmark (Confidentiality focus) & $\times$ \\
\hline
\cite{chatzimiltis2025xaillm} & XAI–LLM: anomaly detection in UEs behaviour using KPMs analysing, enhancing semantic interpretability of telemetry without enforcing integrity. & $\times$ & \checkmark (Semantic interpretation) & $\times$ \\
\hline
\cite{bib-15}, \cite{bib-19} & Experimental study of adversarial attacks (FGSM, PGD) on ML-based xApps and adversarial training for robustness improvement. & $\times$ & \checkmark (Attack analysis / robustness) & $\times$ \\
\hline
\cite{10613612} & Interface-level security using IPsec encryption and Autoencoder defences during training. & \checkmark (Encrypted channel) & \checkmark (Training-phase mitigation) & $\times$ \\
\hline
\cite{10.1145/3643833.3656118} & µONOS RIC penetration testing revealing authentication and API misconfigurations. & \checkmark (API integrity) & $\times$ & $\times$ \\
\hline
\cite{bib-12} & Experimental evaluation of E2-induced message delays and losses impacting xApp operations. & \checkmark (Performance-level analysis) & $\times$ & $\times$ \\
\hline
\textbf{This work} & Multi-layer runtime safeguarding: E2 message inspection, telemetry anomaly detection, and xApp attestation integrated within near-RT RIC. & \textbf{\checkmark (Runtime validation)} & \textbf{\checkmark (Semantic / temporal validation)} & \textbf{\checkmark (Runtime attestation)} \\
\hline
\end{tabular}
\end{table*}

\section{System Architecture}
\label{sec:architecture}

This section presents the architecture of our proposed runtime security framework for the near-RT RIC in Open RAN. The framework integrates lightweight, modular safeguards into the near-RT control loop to detect and mitigate runtime threats targeting E2 messages, telemetry data, and xApp execution. Although this work demonstrates the framework on FlexRIC, the design is intentionally platform-agnostic: all safeguarding modules operate at standardised RAN–RIC interaction points, allowing the same principles to be adopted across other near-RT RIC implementations. A detailed discussion of portability considerations is provided in Section~\ref{sec:discussion}.\ref{sec:general}.

In addition to implementing these safeguarding components, the framework provides a systematic guideline for mapping detection and mitigation mechanisms to specific threat categories. By defining a structured threat classification and placement strategy, it enables operators to determine optimal deployment locations for each detection module within the control loop, while mitigation units are consistently placed within the near-RT RIC platform.

Although the guideline focuses on the principal runtime threats addressed in this paper, it is designed to be extensible, supporting future integration of additional safeguards as new threat vectors emerge. This approach enhances the adaptability and scalability of near-RT RIC security architectures with a clear methodology for deploying effective, layered runtime protections, beyond the solutions implemented in this work.

\subsection{Threat Categorisation Framework and Implementation Mapping}

Our runtime security framework targets operational-time threats to the near-RT RIC, which can exploit externally influenced inputs or components to manipulate real-time control decisions if left unmitigated. To systematically address these threats, we first define a general categorisation framework that classifies threats in near-RT control loop based on their nature and impact, and guides the placement of detection and mitigation modules within the near-RT control loop.

Table~\ref{tab:threat_guideline} presents this guideline, categorising threats into three levels, Message-Level, Data-Level, and Control Logic-Level, and recommending corresponding detection and mitigation placements. Building on this guideline, Table~\ref{tab:implemented_modules} summarises the specific threat types addressed in this study, providing representative attack examples, the implemented detection modules, and their placement within the near-RT RIC control loop. These implemented safeguards demonstrate practical instantiations of the general framework and validate its effectiveness against key runtime threat vectors.

\setlength{\tabcolsep}{4pt}
\renewcommand{\arraystretch}{1.2}
\renewcommand{\tabularxcolumn}[1]{>{\raggedright\arraybackslash}m{#1}}

\begin{table*}[t]
\caption{General guideline for categorising runtime threats in near-RT RIC and recommended detection/mitigation placement.}
\label{tab:threat_guideline}
\centering

\begin{tabularx}{\textwidth}{|m{2.5cm}|m{3cm}|m{3.6cm}|X|m{4cm}|}
\hline
\textbf{Threat Category} & \textbf{Threat Root} & \textbf{Attack Target} & \textbf{Detection Module Placement} & \textbf{Mitigation Module Placement} \\
\hline
Message-Level & Compromised E2 nodes or E2 interface & Near-RT RIC messaging ingress (e.g. protocol messages) & E2 terminal (message parsing/inspection stage) & Near-RT RIC platform \\
\hline
Data-Level & Compromised E2 nodes, E2 interface, or xApps injecting falsified data & Telemetry data and inference inputs used by xApps & Trusted xApp or function within near-RT RIC platform & Near-RT RIC platform \\
\hline
Control Logic-Level & Tampered xApp or malicious xApp logic & Near-RT RIC control decisions issued by xApps & Function within near-RT RIC platform (attestation engine) & Near-RT RIC platform \\
\hline
\end{tabularx}
\end{table*}

\setlength{\tabcolsep}{4pt}
\renewcommand{\arraystretch}{1.2}
\renewcommand{\tabularxcolumn}[1]{>{\raggedright\arraybackslash}m{#1}}

\begin{table*}[t]
\caption{Implemented detection modules for each threat category.}
\label{tab:implemented_modules}
\centering

\begin{tabularx}{\textwidth}{|m{2.4cm}|m{4.2cm}|m{4cm}|X|m{3cm}|}
\hline
\textbf{Threat Category} & \textbf{Example Attack} & \textbf{Implemented Detection Module} & \textbf{Detection Placement} & \textbf{Mitigation Placement} \\
\hline
Message-Level & Injection of syntactically valid but semantically malicious E2 messages to trigger misbehaviour & E2 Message Inspector Module (signature-based detection) & E2 terminal (ingress point) & Near-RT RIC platform \\
\hline
Data-Level & Poisoning of KPM telemetry reports to mislead AI/ML-based xApps & KPM Poisoning Detection xApp (LSTM-based anomaly detection) & Near-RT RIC platform (as standalone xApp) & Near-RT RIC platform \\
\hline
Control Logic-Level & Tampering with xApp binaries to issue unauthorised control actions & Runtime xApp Attestation Module (hash-based integrity verification) & Near-RT RIC platform (attestation engine) & Near-RT RIC platform \\
\hline
\end{tabularx}
\end{table*}

\subsection{Architectural Overview and Deployment Strategy}

\begin{figure}[t]
    \centering
    \includegraphics[width=1\linewidth]{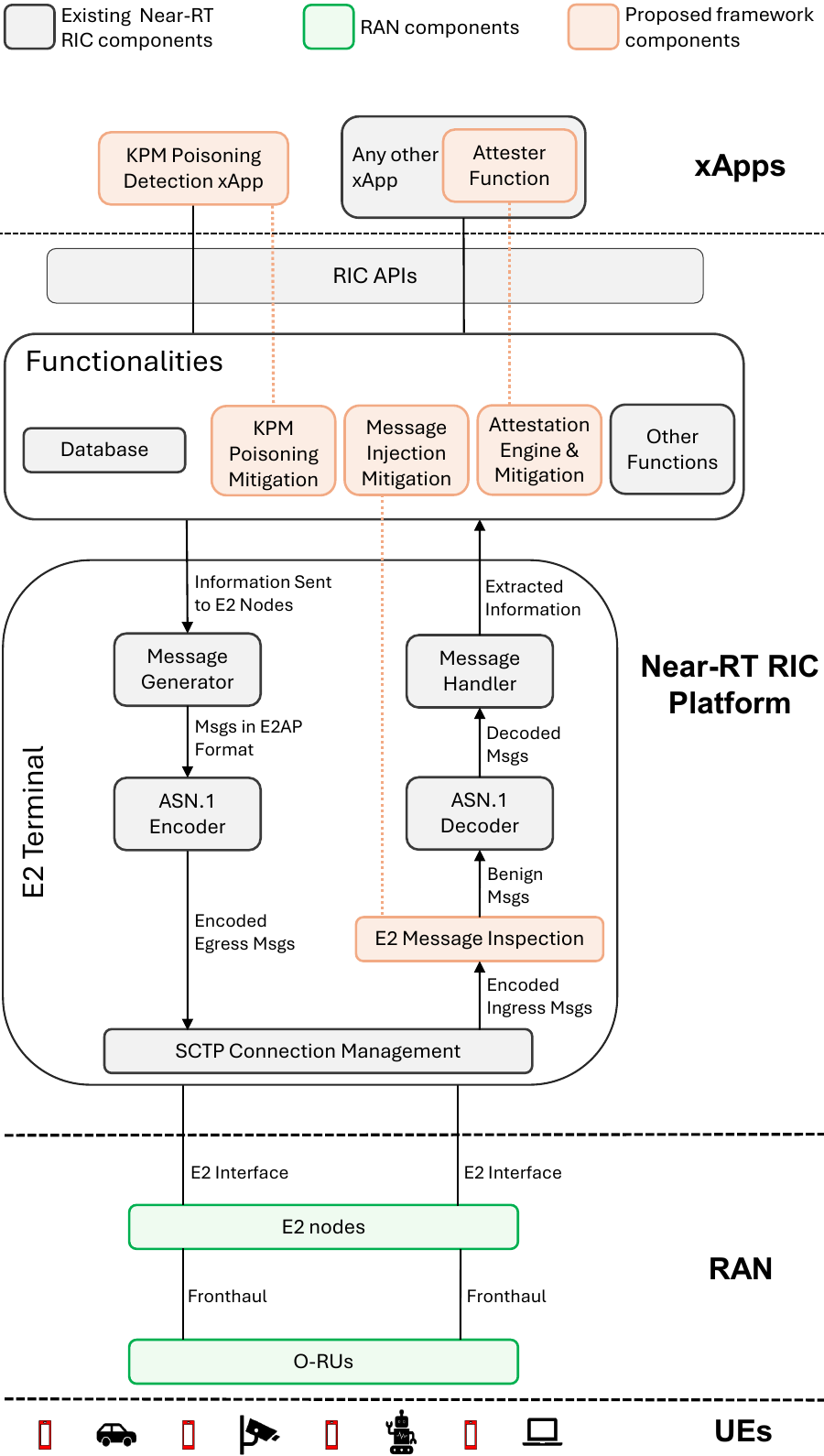}
    \caption{Architecture of the proposed runtime security framework for the near-RT RIC.}
    \label{fig:placement}
\end{figure}

Figure~\ref{fig:placement} illustrates how the proposed runtime security components are integrated within the near-RT RIC architecture. Each component is positioned to intercept or validate a specific stage of the control loop, enabling early threat detection and mitigation with minimal disruption to existing RIC functions. A mitigation unit is associated with each detection component, all of which are placed at the near-RT RIC platform. The components integrate into the standard RIC workflow and do not require changes to the RIC messaging pipeline. The interaction between xApps and the near-RT RIC platform is performed through RIC platform APIs. Where necessary, additional internal service APIs can be introduced—for example, to support runtime xApp attestation—without modifying the external behaviour or deployment model of xApps.

The \textit{E2 message inspection module} operates at the ingress point of the E2 interface. It monitors incoming messages and applies signature-based filtering to detect syntactically correct but semantically malicious patterns, and redirects malicious messages to the mitigation unit. By inspecting control and configuration messages in real-time, it helps prevent adversarial E2 nodes or MitM attackers from injecting harmful signalling. This module is deployed internally within the near-RT RIC as a sidecar or plugin to the E2 terminal, enabling real-time inspection prior to message dispatch.

The \textit{KPM poisoning detector} is implemented as a standalone xApp. It subscribes to telemetry streams such as KPM reports and uses anomaly detection to flag statistical deviations or behavioural inconsistencies that may indicate poisoning. This component works with a mitigation unit to protect AI/ML-driven xApps from making control decisions based on falsified input data. It is deployed via standard xApp lifecycle management procedures and uses E2SM-based subscriptions to access telemetry in a non-intrusive manner.

The \textit{runtime xApp attestation module} is distributed across two components: (i) an \textit{attestation engine} embedded within the near-RT RIC, and (ii) an \textit{attester function} integrated into each xApp that is subject to verification. The xApp computes its runtime hash and reports it to the attestation engine, which validates it against a trusted registry. The associated mitigation unit takes an appropriate action once an integrity violation is detected. This split design enables decentralised attestation while retaining centralised verification within the platform.

Minimal inter-component coordination is required, enabling lightweight and parallel execution of the safeguards. Alerts and detection outputs from individual modules can optionally be routed to a central logging or policy engine in SMO to support broader security workflows and auditability. Furthermore, the modular nature of the framework supports incremental adoption, allowing operators to deploy only the components that align with their threat models and resource constraints.

Together, these components form a layered defence strategy across the control loop, covering input validation, inference integrity, and xApp trustworthiness.

\section{Integrated Runtime Defence Components}
\label{sec:components}
As outlined above, our framework comprises three detection modules, each paired with a corresponding mitigation unit deployed within the near-RT RIC platform. These components are designed to operate independently, enabling flexible deployment, low overhead, and seamless integration into existing workflows. A key feature of our approach is the use of \textit{policy-driven mitigation}, which allows responses to detected threats to be dynamically tailored based on threat severity, source, and operator preferences. The mitigation logic is designed to be extensible, enabling operators to define, refine, or remove mitigation actions over time without modifying the detection modules. This section provides a detailed description of each component’s architecture, its operational role within the framework, and how it contributes to the overall goal of securing the near-RT RIC against runtime threats.

\subsection{Signature-Based E2 Message Inspection}
\label{sec:E2 message inspection}

The E2 interface plays a critical role in Open RAN systems, enabling communication between the near-RT RIC and underlying RAN components such as O-CUs and O-DUs. Positioned at the ingress of the control loop, the E2 interface represents an attractive target for adversaries seeking to disrupt or manipulate real-time network behaviour. Ensuring its integrity is therefore essential to maintaining secure and trustworthy RAN operation.

A notable attack vector involves the injection of crafted patterns—commonly referred to as \textit{signatures}—into syntactically valid E2 messages. These signatures are not necessarily harmful in terms of payload content but can serve different purposes. Some may act as benign indicators of known threats, used for testing or simulation, while others are designed to exploit weaknesses in downstream logic. Our focus is on the latter class: signatures that, although structurally compliant, can provoke unintended or unauthorised control actions when processed by vulnerable RIC components. These may be embedded within permissible fields such as \texttt{OCTET STRING} or \texttt{PrintableString} in protocols like E2AP, E2SM-KPM, E2SM-RC, or E2SM-CCC. While often overlooked by basic validation mechanisms, such signatures can be used to trigger control misbehaviours or bypass safeguards in poorly designed systems.

\begin{figure}[t]
    \centering
    \includegraphics[width=1\linewidth]{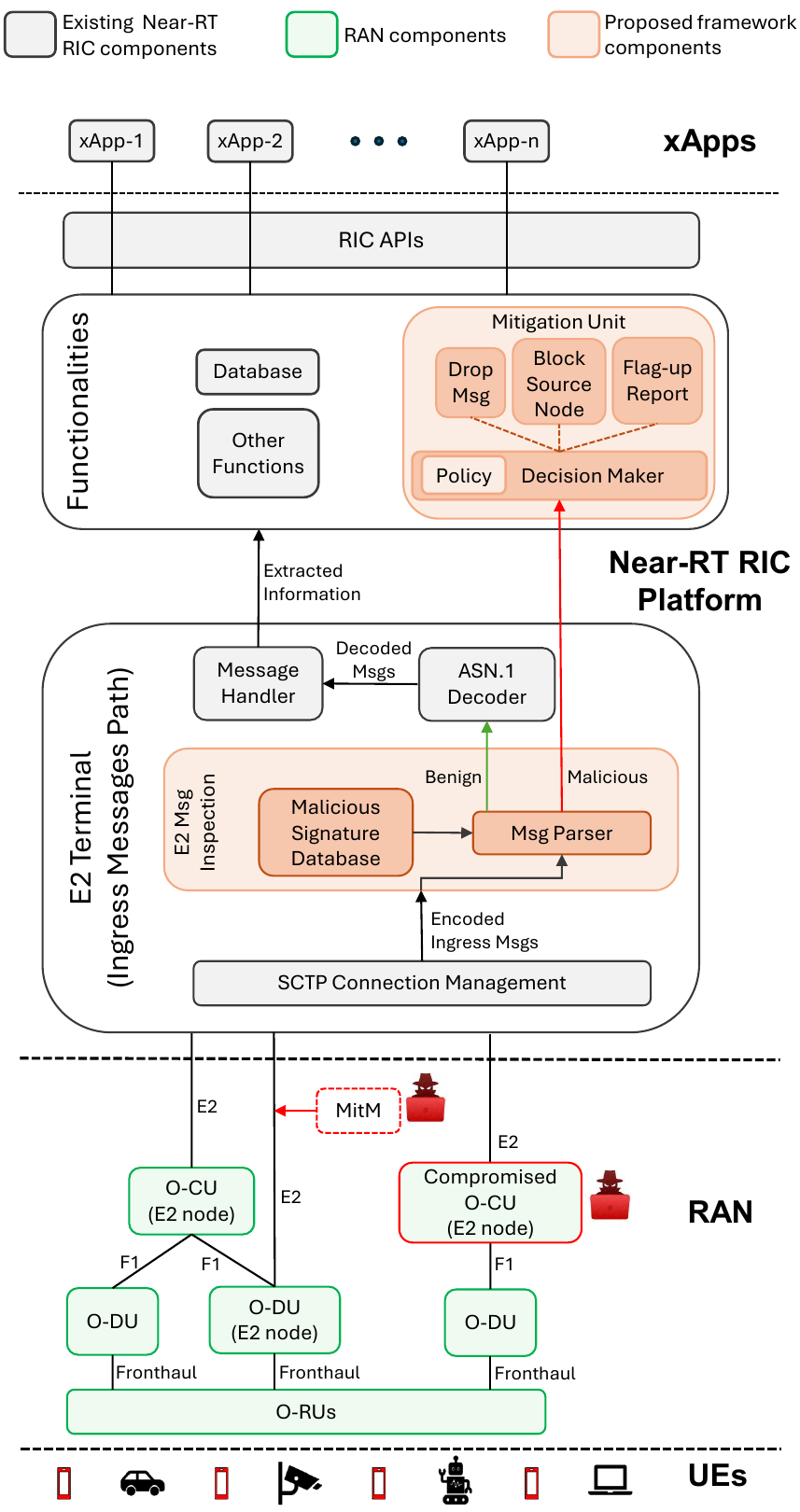}
    \caption{E2 message inspection and associated mitigation.}
    \label{fig:e2inspector}
\end{figure}

To counter this threat, we implement a signature-based detection module integrated at the ingress of the near-RT RIC. Figure~\ref{fig:e2inspector} illustrates the architecture of this component, its associated mitigation unit, and potential threat roots. Operating ahead of the message processing pipeline at the E2 terminal, the module scans all inbound E2 messages for known malicious patterns. It is done before the ASN.1 decoder, as the inspector treats the payload as a string and the patterns are detectable at this stage. If a match is detected, the message is directed to the mitigation unit. This unit decides which action(s) to be taken based on its policy that determines the action(s) for each kind of signatures. The policy is determined by the operator. If a message contains more than one signature, the mitigation unit takes the union of the action sets associated with those signatures. It can choose one or more of the possible actions: (i) \textit{dropping the message}, (ii) \textit{blocking the source E2 node}, (iii) \textit{reporting the attack to the administration body that can be a human or a function in SMO}. Benign messages are sent to the normal message handling path. This procedure prevents malicious messages from influencing xApps or other internal control logic.

The current implementation employs a naive pattern search algorithm, chosen for its efficiency given the short and structured nature of typical E2 messages. Our measurements show that most E2 messages are under a few hundred bytes, with the exception of E2 Setup Requests ($\approx 25$ KB). These Setup messages are inspected like all other messages; however, they occur only during E2 node (re)initialisation and therefore do not appear within the periodic near-RT control loop. As a result, the latency of inspecting these messages does not affect near-RT timing constraints in the same way as RIC Indication messages, which are exchanged frequently, e.g. every 1 second, and directly influence control decisions. We report their inspection latency in Table \ref{tab:e2_latency}, and even in the worst case it remains well within acceptable limits for operations outside the near-RT loop. In cloud-native deployments, events such as scaling or failover may cause Setup Requests to occur more frequently. Even in such scenarios, these messages still lie outside the near-RT control pathway, meaning that their inspection latency does not affect near-RT timing constraints. With a rule set of signatures corresponding to 100 known attacks, the naive search achieves sufficient performance. However, should the signature set expand or messages created in large sizes, scalable approaches such as the Aho-Corasick algorithm or trie-based matching could be considered.

This inspection mechanism complements broader runtime safeguards by providing a first line of defence at the near-RT RIC RAN-side entry point. It reinforces the trust boundary of the near-RT RIC by preemptively filtering suspicious inputs before they can affect system behaviour.

\subsection{KPM Poisoning Detection}
\label{ssec:kpm-poisoning}

Key Performance Measurements (KPMs) are critical inputs to control functions in Open RAN, directly influencing decisions made by the near-RT RIC and its xApps. These reports guide operations such as traffic steering, mobility management, and energy-efficiency optimisation. However, their central role also makes them a high-value target for adversaries and should therefore be subject to anomaly detection, an established practice in network security and monitoring \cite{KAYA2025111372} but not well-explored in Open RAN context. In a KPM poisoning attack, falsified measurement reports are injected into the control loop, misleading the system and potentially degrading performance or violating service-level objectives.

In our earlier work~\cite{11028721}, we introduced and analysed this threat in detail, demonstrating that even partial telemetry poisoning can disrupt different use cases with varying severity. To address this, we proposed a detection approach based on sequence-learning neural networks, capable of capturing temporal dependencies in KPM telemetry. Although our initial prototype employed Long Short-Term Memory (LSTM) networks due to their effectiveness in modelling sequential data, in this work we generalise the approach and explore a broader family of time-series classification models. Specifically, we evaluate recurrent architectures (LSTM and GRU), convolutional sequence models (1D-CNN), and attention-based encoders (Transformer), allowing us to assess how different temporal modelling paradigms behave under telemetry poisoning. The final detection model, irrespective of the underlying neural architecture, can be integrated into our broader runtime security framework for the near-RT RIC. The detector processes sequences of KPM records, where each record corresponds to a specific user and timestamp, and includes multiple KPMs. By learning temporal patterns and detecting deviations indicative of tampering, the model provides an early warning of poisoning attempts before the data is consumed by control xApps.

In our previous work, the poisoning detector was proposed as a module integrated within the near-RT RIC core to conceptually demonstrate tight integration. However, for practical deployment, some components of the framework, such as the E2 message inspector and mitigation units, remain platform-integrated due to their nature and privilege requirements, while we implemented the KPM poisoning detector as a standalone xApp. This design maintains modularity, enables deployment flexibility, and allows operators to selectively route xApps through the detector based on their latency tolerance. For example, ultra time-sensitive xApps may access KPM reports directly to preserve strict timing requirements, while other xApps can benefit from filtered and verified telemetry inputs with minimal deployment complexity.

The poisoning detector xApp subscribes to E2 nodes, collects KPM reports, and stores them in a shared database. Other xApps consume these stored KPMs rather than retrieving them directly from E2 nodes, allowing them to operate on telemetry that has already been verified for integrity. This indirect access is suitable for the majority of xApps, as KPM reporting is typically configured at a one-second interval, which aligns with the operational timescale of near-RT RIC control policies in current Open RAN deployments. Use cases requiring sub-second telemetry generally correspond to collaboration with fast control loops implemented at the O-DU (e.g., MAC scheduling or HARQ adaptation). In such cases, direct access to raw KPM reports remains available to preserve performance for ultra time-sensitive operations, while higher-level control logic continues to benefit from the verified telemetry path. An xApp’s choice between consuming verified telemetry or accessing raw KPM reports is governed by operator policy, which can be enforced through subscription filtering or by selectively permitting database bypass for ultra time-sensitive workflows.

As mentioned before, our threat model assumes that the near-RT RIC itself is a trusted entity, and that poisoning occurs prior to the ingestion of telemetry. This includes two primary vectors: (i) MitM attacks on the E2 interface, where telemetry is intercepted and altered in transit, and (ii) compromised E2 nodes, where adversaries directly control the data source. While our prior work~\cite{11028721} focused on characterising the attack and evaluating detection under emulated conditions, this paper embeds the defence mechanism into a deployable xApp, forming a key component of the proposed runtime protection framework for Open RAN. This xApp is accompanied by a mitigation unit at the near-RT RIC platform. Once the xApp detects a poisoned data, it alerts the mitigation unit, which then decides on the action(s) should be taken based on the policy. The policy can be adjusted based on the operator's preferences. One example policy can be taking actions based on the deviation magnitude and the source E2 node. The deviation magnitude is classified to three levels, small, moderate, and significant. Three actions have been provided: (i) \textit{dropping the poisoned data} for all magnitude classes, (ii) \textit{blocking the source E2 node} for significant magnitude class, and (iii) \textit{reporting the issue to the administration body} for moderate and significant magnitude classes.

\begin{figure}[t]
    \centering
    \includegraphics[width=1\linewidth]{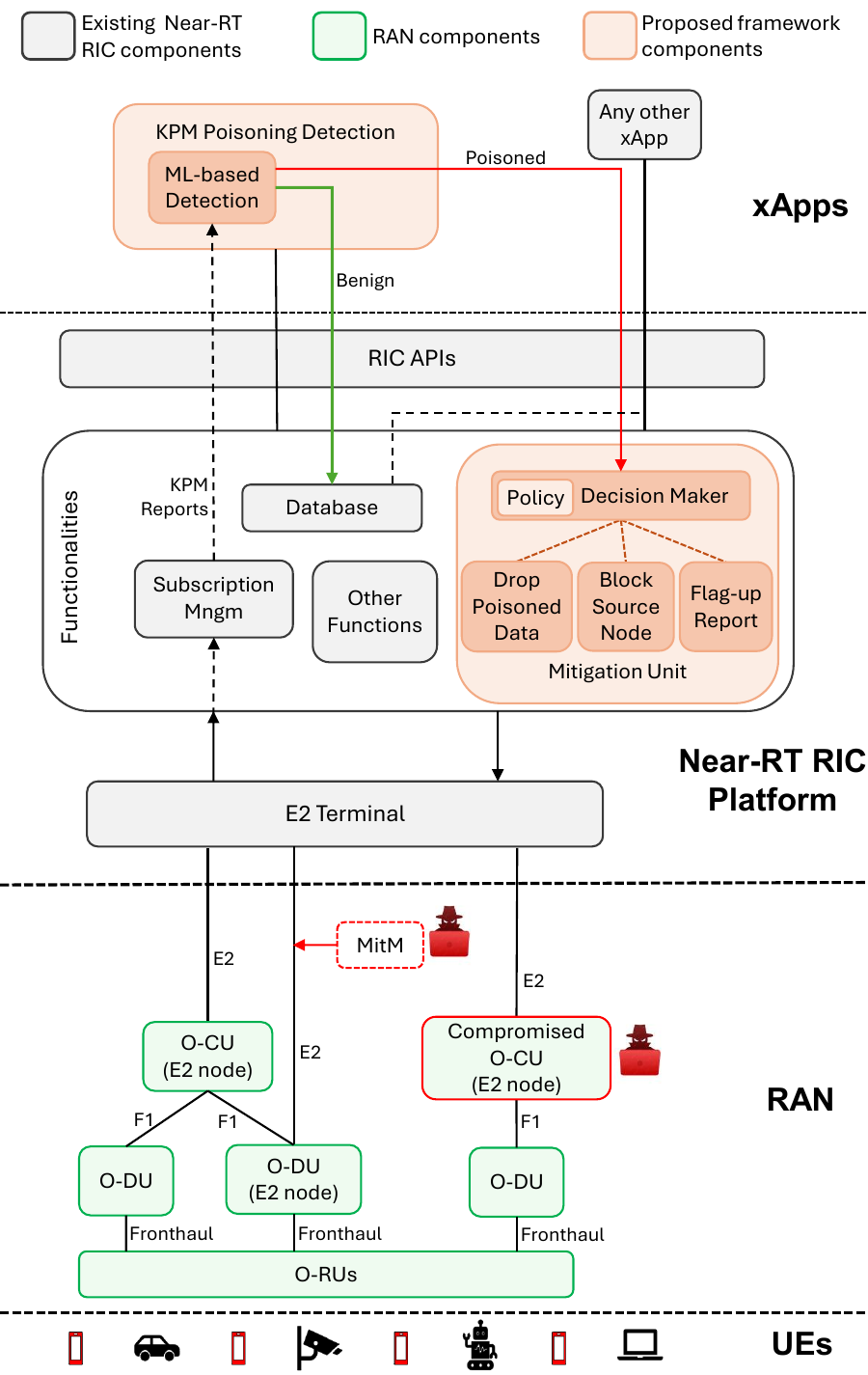}
    \caption{KPM poisoning detection and mitigation.}
    \label{fig:kpm}
\end{figure}

Figure~\ref{fig:kpm} shows the architecture of the solution and the roots of the threat. In summary, the KPM poisoning detector plays a critical role in safeguarding the inference inputs of control xApps. By combining temporal sequence learning with flexible, policy-driven mitigation, it provides a robust and adaptive line of defence against telemetry-based attacks. As a standalone xApp integrated within our modular framework, it enables scalable, operator-configurable protection without disrupting existing control-loop operations.

\subsection{Runtime xApp Attestation}
\label{sec:xappattestation}

The open and modular design of the near-RT RIC enables third-party developers to deploy diverse xApps that dynamically interact with RAN functions and system resources. While this promotes innovation and adaptability, it also introduces significant security risks. In particular, a compromised or malicious xApp may abuse its operational privileges, issue unauthorised control actions, or leak sensitive data. Static onboarding validation is insufficient for maintaining long-term trust, highlighting the need for a runtime attestation mechanism that can continuously verify the integrity of executing xApps.

\begin{figure}[t]
    \centering
    \includegraphics[width=1\linewidth]{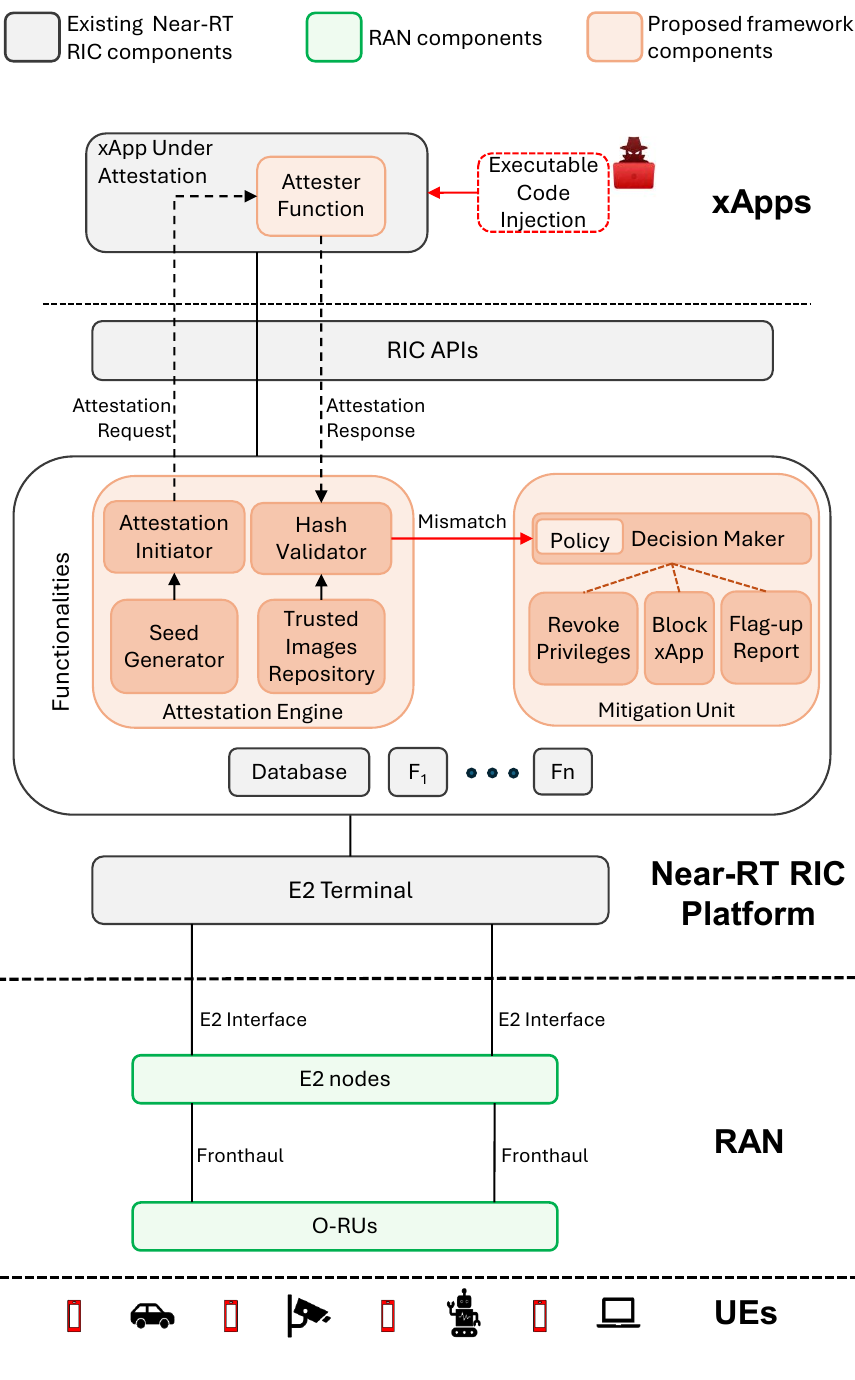}
    \caption{xApp attestation and integrity violation mitigation.}
    \label{fig:attest}
\end{figure}

Figure~\ref{fig:attest} shows our attestation module and its associated mitigation unit components and architecture. The attestation mechanism consists of two coordinated components: (i) an \textit{attestation engine} deployed within the near-RT RIC, and (ii) a lightweight \textit{attester function} embedded in each xApp that is to be verified. The RIC-side engine initiates periodic attestation sessions and validates the integrity proofs, while each xApp's attester is responsible for computing responses to these challenges. Prior to deployment, each xApp is provisioned with a trusted snapshot of its binary memory layout.

During runtime, the attestation engine issues a challenge comprising a fresh random seed. The xApp computes a hash over its active memory region using this seed and returns the result. The RIC engine then independently computes the expected hash using the same seed and the reference memory image. If the two hashes match, the xApp is deemed unmodified; a mismatch indicates potential code injection or unauthorised modification. An integrity violation alert is sent to the mitigation unit. Three possible mitigation actions are: (i) \textit{revoking the xApp’s control privileges}, (ii) \textit{blocking the xApp}, and (iii) \textit{reporting the incident to the administrative entity}. The policy is that action (iii) is always taken, as any detected violation must be recorded and addressed by network management, either via automated systems or human oversight. Actions (i) and (ii) may also be enforced based on the xApp type and its assigned privilege level. For instance, high-impact xApps responsible for mobility management, traffic steering, or security enforcement are more likely to be immediately blocked or revoked upon a confirmed violation to prevent harm to network operation. In contrast, lower-priority or read-only xApps may continue to operate in a restricted mode while further investigation is carried out. This tiered response ensures that mitigation is proportionate to potential risk, while preserving service continuity where possible.

The attestation process is designed to be lightweight, scalable, and non-intrusive. It can be executed frequently with minimal performance overhead, as each xApp is attested independently and the mechanism does not interfere with near-RT control loop timing. The use of seed-based hashing prevents replay attacks and ensures response freshness. This design supports secure and continuous integrity monitoring in dynamic RAN environments.

Additionally, there exists a fundamental trade-off between the level of security provided by attestation and the associated performance cost. To balance this, the frequency of attestation can be configured based on the type and criticality of each xApp. For example, xApps with higher control privileges or greater impact on RAN behaviour—such as mobility or resource allocation controllers—may warrant more frequent integrity checks, whereas monitoring xApps may require less stringent validation. In this paper, we delegate the responsibility of defining attestation intervals to the network operator, allowing customisation based on operational risk profiles and resource constraints.

By enabling continuous runtime validation of xApps, this mechanism enhances the trustworthiness of the near-RT RIC control loop and mitigates risks stemming from dynamic code tampering or unauthorised logic injection.

\subsection{Summary of Runtime Behaviour}
\begin{figure}[t]
    \centering
    \includegraphics[width=1\linewidth]{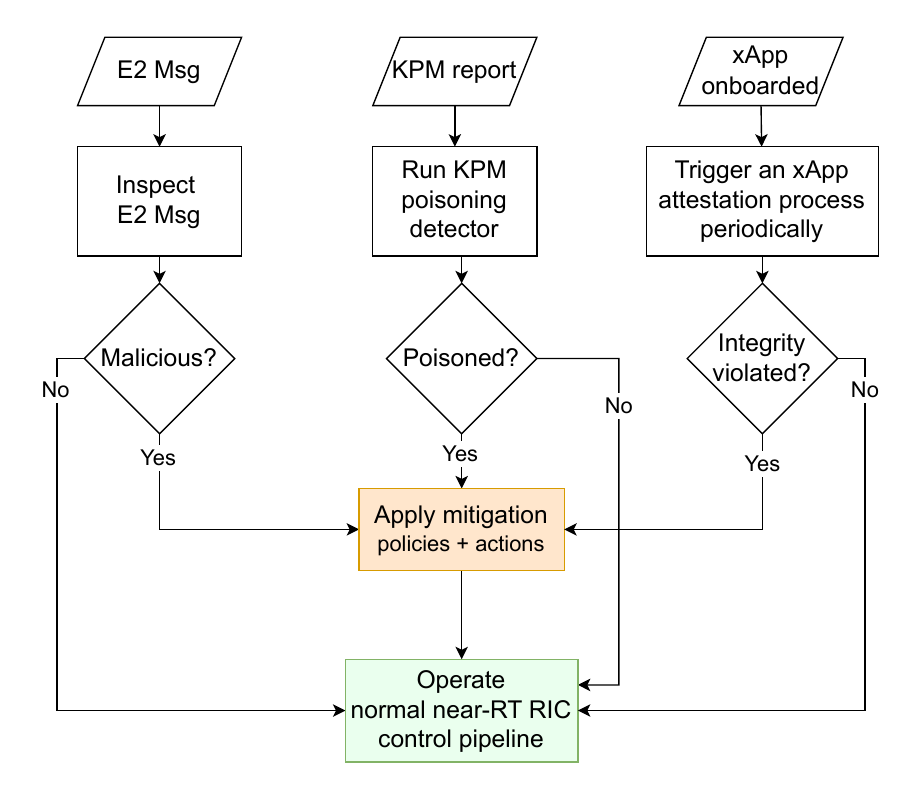}
    \caption{End-to-end runtime workflow.}
    \label{fig:runtime_flow}
\end{figure}

To complement the component-level descriptions above, Table~\ref{tab:components_summary} summarises the runtime operation of the three proposed safeguards by consolidating each module’s trigger conditions, required inputs, core processing steps, and resulting actions. While the table captures the behaviour of each component in isolation, it is equally important to illustrate how these components operate together within the near-RT RIC execution loop. Accordingly, Figure~\ref{fig:runtime_flow} provides an end-to-end flowchart of the proposed multi-layer defence framework. The diagram shows how incoming E2 messages, periodic KPM reports, and xApp lifecycle events are directed through three parallel analysis paths corresponding to message-level inspection, data-level poisoning detection, and control-logic attestation. Each path applies its respective detection logic and, upon identifying a threat, invokes the mitigation stage, whereas benign inputs are forwarded to the standard near-RT RIC control pipeline. Together, this workflow highlights how the three safeguards collectively reinforce the trustworthiness and resilience of the near-RT RIC’s runtime decision-making process.

\begin{table*}[t]
\centering
\footnotesize
\caption{Runtime operation of the proposed security components.}
\label{tab:components_summary}
\begin{tabular}{|p{3.2cm}|p{2.5cm}|p{2.5cm}|p{4cm}|p{4cm}|}
\hline
\textbf{Module} & \textbf{Trigger / Frequency} & \textbf{Input} & \textbf{Operation Performed} & \textbf{Output / Action} \\
\hline

E2 Message Inspector &
Per E2 message &
Raw E2AP PDUs &
Signature search to detect known malicious signalling patterns &
Flag suspicious message; invoke mitigation action \\
\hline

KPM Poisoning Detector &
Per reporting interval &
Periodic KPM reports / telemetry &
ML-based anomaly scoring to detect poisoning-like deviations &
Raise alert; suppress or isolate anomalous telemetry \\
\hline

xApp Attestation Module &
Periodic &
xApp executable and reference image &
Hash computation and comparison to verify integrity &
Block, restart, or prevent execution of untrusted xApps \\
\hline

\end{tabular}
\end{table*}

\section{Implementation Details and Experimental Results}
\label{sec:evaluation}

The proposed runtime defence framework was implemented and evaluated using a testbed comprising FlexRIC as the near-RT RIC platform and a commercial emulator for RAN emulation. FlexRIC and xApps were deployed on a dedicated server acting as the near-RT RIC, while the emulator was deployed on a separate server to emulate E2 nodes and user equipment (UEs). In this setup, each E2 node established an E2 interface connection to FlexRIC. All detection and mitigation components were integrated within FlexRIC. Table~\ref{tab:implement} summarises the implementation details, overhead sources, and deployment notes for the three defence components integrated into the testbed. Both servers had identical specifications: Intel(R) Xeon(R) CPU E5-2667 v2 @ 3.30GHz, 192~GB RAM, running Ubuntu~24 (64-bit).

This setup enabled realistic emulation of E2 node behaviour and traffic, with FlexRIC executing the detection and mitigation components in a near-real-time environment while the emulator generated dynamic RAN scenarios and telemetry for evaluation. 

Since no existing multi-layer runtime defence framework directly targets the combined threats of message-level, data-level, and control-logic-level attacks within the near-RT RIC, the only meaningful baseline comparison in our experiments was against the unprotected system (FlexRIC + emulator). This comparison, presented in Section~\ref{sec:evaluation}.\ref{testall}, demonstrates the incremental overhead introduced by each of our proposed safeguarding components. Due to the absence of similar deployable solutions in the literature, our focus remains on evaluating the effectiveness and practical integration of our framework within a realistic testbed environment.

In the following subsections, experimental scenarios and results are presented and discussed. First, the experiments for each of the three defence modules are described, followed by a use-case evaluation with all modules onboard. Each scenario was repeated over ten runs to ensure statistical reliability.

\begin{table*}[t]
\centering
\footnotesize

\caption{Implementation details of the runtime defence components.}
\label{tab:implement}
\begin{tabular}{|p{3cm}|p{5.3cm}|p{4cm}|p{4.1cm}|}
\hline
\textbf{Module} & \textbf{Platform / Tools} & \textbf{Overhead Source} & \textbf{Notes} \\
\hline

E2 Message Inspector &
C implementation within FlexRIC environment library &
Per-message signature search &
Lightweight processing due to small E2AP message sizes \\
\hline

KPM Poisoning Detector &
xApp implemented in Python + C for detection, added C functions for mitigation &
ML inference per reporting interval &
Runs as separate monitoring process; relaxes near-RT timing impact \\
\hline

xApp Attestation Module &
Integrated SHA-256 hashing + messaging modules in both FlexRIC and xApp sides implemented in C &
Hash computation and messaging at each attestation process &
No direct effect on control-loop runtime latency \\
\hline

\end{tabular}
\end{table*}

\subsection{E2 Message Inspector}

To evaluate the E2 message inspector, a scenario with four E2 nodes, twelve cells, and varying numbers of UEs per cell was implemented on the RAN emulator. The emulator's signature injection feature injected signatures corresponding to 100 known attacks from the CVE database. The percentage of malicious nodes was set to 50\%, resulting in two benign and two malicious E2 nodes. Additionally, the percentage of injected malicious E2 messages was set to 50\%, as a 100\% injection rate would prevent malicious E2 nodes from establishing E2 connections to FlexRIC by blocking all E2 setup requests.

On the FlexRIC side, the inspector module was deployed on the E2 terminal, with its mitigation module implemented as an internal functionality. A KPM monitoring xApp was used to trigger different types of E2 messages from E2 nodes to FlexRIC. This xApp subscribed to E2SM KPM, receiving reports every second from all E2 nodes, and operated for 100 seconds per run.

The mitigation policy was configured to drop detected malicious messages. Blocking malicious E2 nodes was not applied, as this would terminate message reception within a few seconds, preventing experiment completion. As malicious messages were redirected to the mitigation unit and removed from the near-RT control loop, mitigation execution time was not included in control loop timing evaluations. For the same reason, flag-up report creation time was excluded. The primary evaluation metric was inspection latency, as it is the only part affecting the control loop time. The detection rate remained consistently at 100\% due to the rule-based nature of the inspector.

\begin{figure}[t]
    \centering
    \includegraphics[width=.95\linewidth]{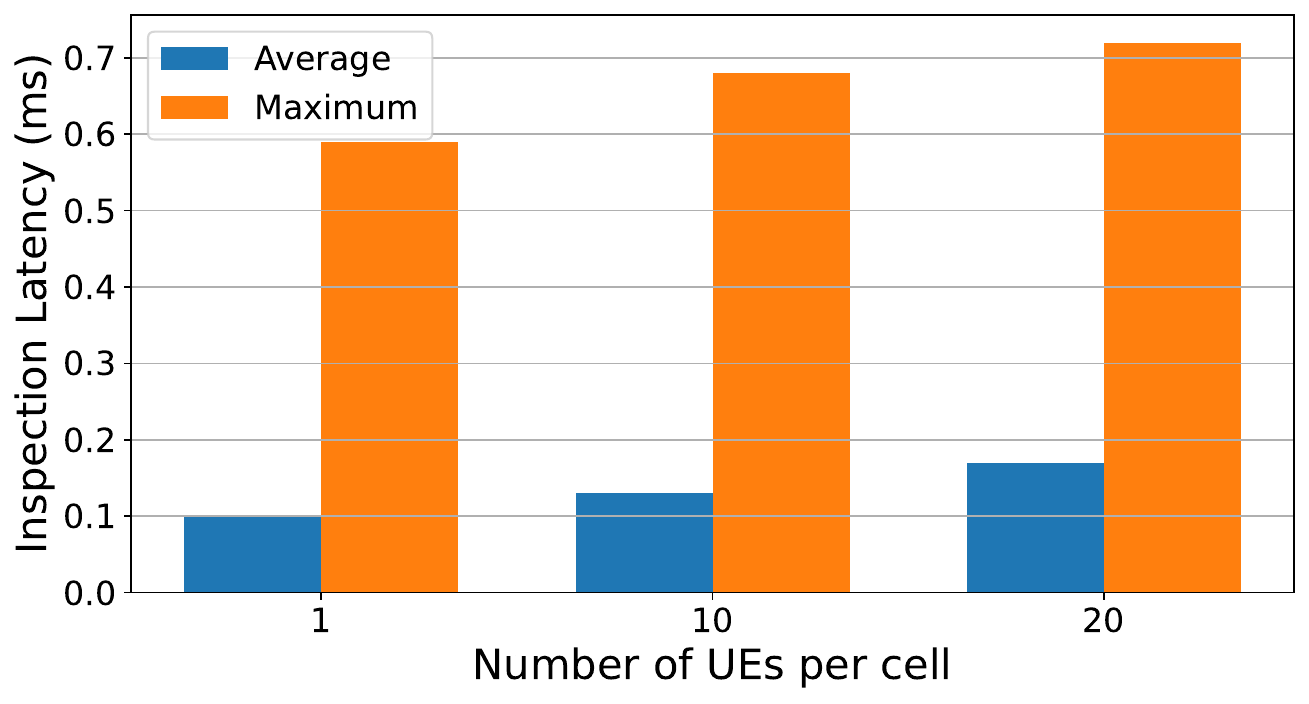}
    \caption{RIC Indication message inspection latency.}
    \label{fig:e2_latency}
\end{figure}

Four types of ingress E2 messages were observed. Table~\ref{tab:e2_sizes} presents their sizes under different user densities. Only RIC Indication messages, containing KPM reports, showed size variation with user numbers. Figure~\ref{fig:e2_latency} shows the end-to-end inspection latency for RIC Indication messages under different UE loads. These messages are the most frequent E2 message type and the only ones whose processing delay directly uses the near-RT control loop timing budget. As the figure illustrates, the inspection latency increases only slightly and approximately linearly with the number of UEs, reflecting the corresponding growth in message volume. Even under higher UE loads, the measured latency remains well within near-RT timing constraints, confirming that the message-level defence component scales gracefully without risking control-loop violations.

\begin{table}[t]
\centering
\caption{E2 message sizes for different user densities.}
\label{tab:e2_sizes}
\footnotesize
\setlength{\tabcolsep}{3pt}
\begin{tabular}{|l|c|c|c|}
\hline
\textbf{Message Type / Size} & \textbf{1 UE/cell} & \textbf{10 UEs/cell} & \textbf{20 UEs/cell} \\
\hline
E2 Setup Request & $\approx$25 KB & $\approx$25 KB & $\approx$25 KB \\
\hline
RIC Subscription Response & 38 B & 38 B & 38 B \\
\hline
RIC Indication & $\approx$100 B & $\approx$150 B & $\approx$200 B \\
\hline
RIC Subscription Delete Response & 22 B & 22 B & 22 B \\
\hline
\end{tabular}
\end{table}

Table~\ref{tab:e2_latency} summarises average and maximum inspection latencies, evaluated with ten UEs per cell. As expected, E2 Setup Request messages, conveying full configuration information, incurred the highest latencies (average 17.22 ms; maximum 66.54 ms). RIC Indication messages, which influence operational time requirements, showed consistently low latencies ($<$1 ms). RIC Subscription Response and Delete Response messages exhibited the fastest inspection times ($<$0.05 ms). Occasional higher maximums were due to instantaneous processing loads and the random placement of malicious signatures within messages by the RAN emulator. Signatures located towards the end of messages took longer to detect due to the naive search approach.

\begin{table}[t]
\centering
\caption{Latency of E2 messages inspector.}
\label{tab:e2_latency}
\begin{tabular}{|l|c|c|}
\hline
\textbf{Message Type} & \textbf{Average (ms)} & \textbf{Maximum (ms)} \\
\hline
E2 Setup Request & 17.22 & 66.54 \\
\hline
RIC Subscription Response & 0.024 & 0.037 \\
\hline
RIC Indication & 0.13 & 0.68 \\
\hline
RIC Subscription Delete Response & 0.017 & 0.044 \\
\hline
\end{tabular}
\end{table}

Overall, inspection latencies did not violate the one-second near-RT control loop constraint. Although some ultra time-sensitive use cases may require even sub-second latencies, the $<$1 ms latency for RIC Indication messages ensures seamless integration without performance degradation.

\subsection{KPM Poisoning Detection}
\label{sec:kpm_poisoning_results}

The KPM poisoning detection module was evaluated as part of its integration into the broader runtime defence framework. As introduced in our earlier work~\cite{11028721}, the detector uses an LSTM-based anomaly detection model to identify temporal deviations in telemetry sequences indicative of poisoning attacks. The detection model processes sequences of KPM records, each containing multiple KPMs for a specific user and timestamp. A sequence length of 10 was used, based on prior analysis showing it provides the best balance between detection accuracy and computational efficiency.

In this implementation, the poisoning detector was integrated into the KPM monitoring xApp within FlexRIC. This design choice, as discussed in Section~\ref{sec:components}.\ref{ssec:kpm-poisoning}, allows operators to integrate detection flexibly and selectively route xApps through the detector based on their latency tolerance.

The evaluation was conducted using the RAN emulator to run a scenario with three base stations, each including an O-CU and three O-DUs (one per cell). Fifty UEs were distributed randomly across the cells. The UEs were evenly split between two network slices, with 25 assigned to the Enhanced Mobile Broadband (eMBB) slice and 25 to the Ultra-Reliable Low Latency Communications (URLLC) slice. For each UE, six key performance measurements (KPMs) were recorded, determined by the emulation platform’s capabilities. Alongside these KPMs, the timestamp and UE identifier were included as input features for the LSTM-based detection model. The specifics of these features are detailed in Table~\ref{tab:KPMs}.

The emulation was first run to collect the dataset and inject anomalies, with 80\% of the data used to train the LSTM model. For live testing, the trained model was deployed within the detection module, and a MitM-based attack model was used to inject anomalies directly into KPM reports streamed from the RAN emulator.

Poisoning attacks were simulated by amplifying the mean and covariance of targeted UEs’ KPM distributions during randomly chosen timeframes, using amplification factors (AF) of 1.2, 1.3, 1.4, and 1.5, as described in~\cite{11028721}. The KPM monitoring xApp subscribed to E2SM KPM reports arriving every 1 second and processed them in real time. Benign records were stored in the database, while malicious records were redirected to the mitigation unit, which dropped poisoned data. Other mitigation policies were not evaluated, as they do not impact the core detection performance metrics presented here.

\begin{table}[t]
\centering
\caption{KPM poisoning dataset features.}
\label{tab:KPMs}
\begin{tabular}{|c|p{2cm}|p{5cm}|}
\hline
\textbf{\#} & \textbf{Feature Name} & \textbf{Description}                                          \\ \hline
1           & Timestamp             & The date and time when the data was recorded.               \\ \hline
2           & UEid                  & The unique identifier for the user equipment.               \\ \hline
3           & UEThpUl               & User equipment uplink throughput.                           \\ \hline
4           & PrbUsedUl             & Number of uplink physical resource blocks used.             \\ \hline
5           & UEThpDl               & User equipment downlink throughput.                         \\ \hline
6           & PrbUsedDl             & Number of downlink physical resource blocks used.           \\ \hline
7           & TotNbrUl\_per\_sec    & Total number of uplink data packets per second.             \\ \hline
8           & TotNbrDl\_per\_sec    & Total number of downlink data packets per second.           \\ \hline
\end{tabular}
\end{table}

As the detection performance of the underlying ML model has already been evaluated in our previous work, the following evaluation metrics are reported here to assess its integration as an xApp and its operational impact:

\begin{itemize}
    \item \textit{Attack Detection Rate (ADR):} Proportion of poisoned records correctly identified.
    \item \textit{False Positive Rate (FPR):} Proportion of benign records incorrectly flagged.
    \item \textit{Latency:} Average end-to-end processing time per KPM record.
\end{itemize}

\begin{figure*}[!t]
    \centering
    \includegraphics[width=\textwidth]{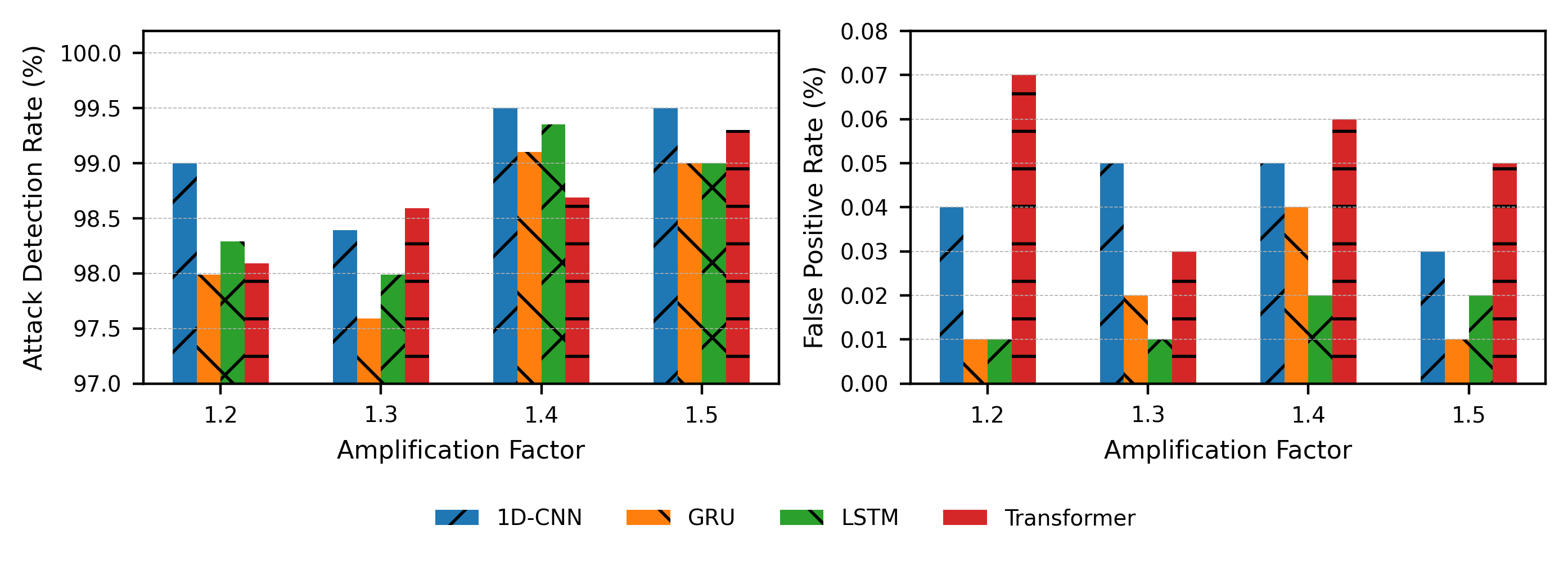} 
    \caption{Comparison of attack detection rate and false positive rate for all evaluated models across amplification factors 1.2--1.5.}
    \label{fig:model_comparison}
\end{figure*}

Table~\ref{tab:kpm_results} summarises the detection performance across all tested amplification factors for the baseline LSTM model. As shown, the detection rate remained consistently high across all attack magnitudes, with minimal false positives. The inference latency remained at 0.15\,ms in all cases, aligning with near-real-time operational requirements for most monitoring and optimisation xApps. 
To complement the LSTM baseline discussed in Section~\ref{sec:components}.\ref{ssec:kpm-poisoning}, we extend the evaluation to several widely used sequence-learning architectures, including GRU, 1D-CNN, and a lightweight Transformer encoder. Table~\ref{tab:comp_results} shows trainable parameters and latencies that are identical across all tested amplification factor values. The relative parameter counts and latencies reflect the architectural properties of the models: LSTMs and GRUs incur sequential computation costs due to recurrent gating, Transformers require attention operations despite having fewer parameters, while 1D-CNNs offer the fastest inference owing to their highly parallel convolutional structure. Figure~\ref{fig:model_comparison} presents a unified comparison across amplification factors 1.2--1.5.
The results show that all architectures achieve consistently high ADR values ($\geq 97.9\%$) with extremely low FPR (typically $\leq0.07\%$), confirming that the poisoning detector remains reliable under varying amplification settings and model used. Among the models, 1D-CNN provided a strong balance of accuracy and computational cost, achieving the lowest inference latency ($\approx 0.05\,\mathrm{ms}$); however, it shows relatively high FPR compared to GRU and LSTM. Finally, we choose LSTM that achieves good enough ADR and latency and the best FPR to use in Sub-Section~\ref{testall} experiments. 1D-CNN with lower latency can be used where the latency is the highest priority. 

In summary, the results demonstrate that the implemented poisoning detector achieves reliable detection performance with low false positive rates and real-time–scale latency, enabling practical deployment within the near-RT RIC environment. Moreover, the comparative study confirms that the defence mechanism is robust across multiple temporal modelling paradigms. Despite architectural differences, all models maintain high detection fidelity, with only marginal variations in ADR and FPR.

\begin{table}[!t]
\centering
\caption{Detection performance metrics of LSTM across different poisoning amplification factors (AF).}
\label{tab:kpm_results}
\begin{tabularx}{\columnwidth}{|c|Y|Y|Y|}
\hline
\textbf{AF} & \textbf{ADR (\%)} & \textbf{FPR (\%)} & \textbf{Latency (ms)} \\
\hline
1.2 & 98.29 & 0.01 & 0.15 \\
\hline
1.3 & 97.99 & 0.01 & 0.15 \\
\hline
1.4 & 99.40 & 0.02 & 0.15 \\
\hline
1.5 & 99.00 & 0.01 & 0.15 \\
\hline
\end{tabularx}
\end{table}

\begin{table}[!t]
    \centering
    \caption{Model size and average inference time per instance.}
    \label{tab:comp_results}
    \begin{tabular}{|l|r|r|}
    \hline
        Model & Trainable Parameters & Latency (ms) \\
        \hline
        LSTM       & 131{,}076 & 0.15 \\
        \hline
        GRU        & 89{,}732  & 0.13 \\
        \hline
        Transformer& 71{,}812  & 0.14 \\
        \hline
        1D-CNN      & 34{,}436  & 0.05 \\
        \hline
    \end{tabular}
\end{table}

\subsection{Runtime xApp Attestation}

The same KPM monitoring xApp used in the E2 message inspector experiments was employed to evaluate the xApp attestation module. The baseline memory footprint of this xApp was approximately 8.5 MB. To assess the impact of xApp size on attestation performance, additional code blocks were inserted to create a larger variant with a footprint of approximately 16 MB, without altering the xApp’s core logic.

For the attestation process, SHA-256 was selected as the hash function due to its strong security guarantees and acceptable computational overhead, providing a sufficiently large output space to minimise collision risks without incurring excessive delays.

To emulate an attack scenario, a code injector was implemented in assembly language to locate the xApp’s memory space in RAM, expand it, and inject arbitrary code. In all experiments, such manipulations were successfully detected by the attestation module. This reliable detection capability is particularly important because integrity violations at the control-logic level cannot be mitigated downstream, and any tampered xApp may directly influence near-RT decision-making. The mitigation policy was configured to block manipulated xApps upon detection; while alternative actions are available, they do not affect the performance metrics evaluated in this study.

The evaluation metric was attestation latency, defined as the elapsed time from when the attestation request was sent by the near-RT RIC to the xApp until completion of the validation process. To allow uninterrupted experiment runs, the attacker code was not executed during latency measurements, as any injection would result in immediate blocking of the xApp. It is worth noting that attestation latency is not part of the near-RT control loop execution time; however it can affect the performance by imposing additional computations. It means that the cost should remain low; however, attestation operates outside the 10 ms–1 s near-RT RIC timing window and does not need satisfy strict near-real-time deadlines. While attestation frequency is operator-defined, results show that latency is sufficiently low to permit attestations every few seconds without significant overhead. In these experiments, attestation was performed once every five seconds.

\begin{figure}[t]
    \centering
    \includegraphics[width=.95\linewidth]{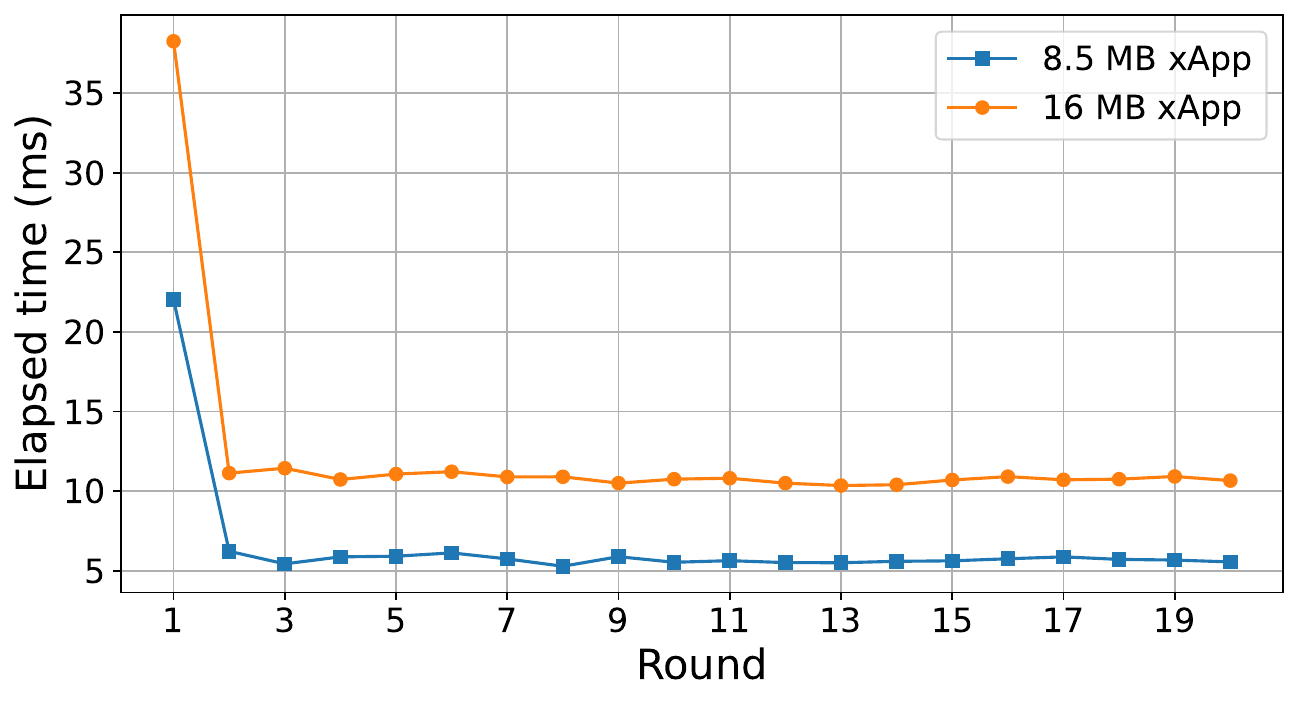}
    \caption{xApp attestation latency.}
    \label{fig:attest_1}
\end{figure}

Figure~\ref{fig:attest_1} shows attestation latency across 20 rounds for both xApp sizes. The first round exhibited higher latencies, approximately 22 ms for the 8.5 MB xApp and 38 ms for the 16 MB xApp, due to cold-start effects from initial disk access to retrieve the trusted reference image. For subsequent rounds, latency stabilised with no significant variation, reflecting the characteristics of SHA-256 as a Merkle–Damgård construction hash function that processes data in fixed-size 512-bit (64-byte) blocks. The low variance across rounds indicates that the hashing and messaging pipeline introduces minimal jitter, an important property for periodic attestation where predictable execution time ensures non-disruptive operation alongside other near-RT RIC tasks. As expected, attestation time was roughly proportional to xApp size, with hashing being the dominant contributor to latency.

\begin{figure}[t]
    \centering
    \includegraphics[width=.95\linewidth]{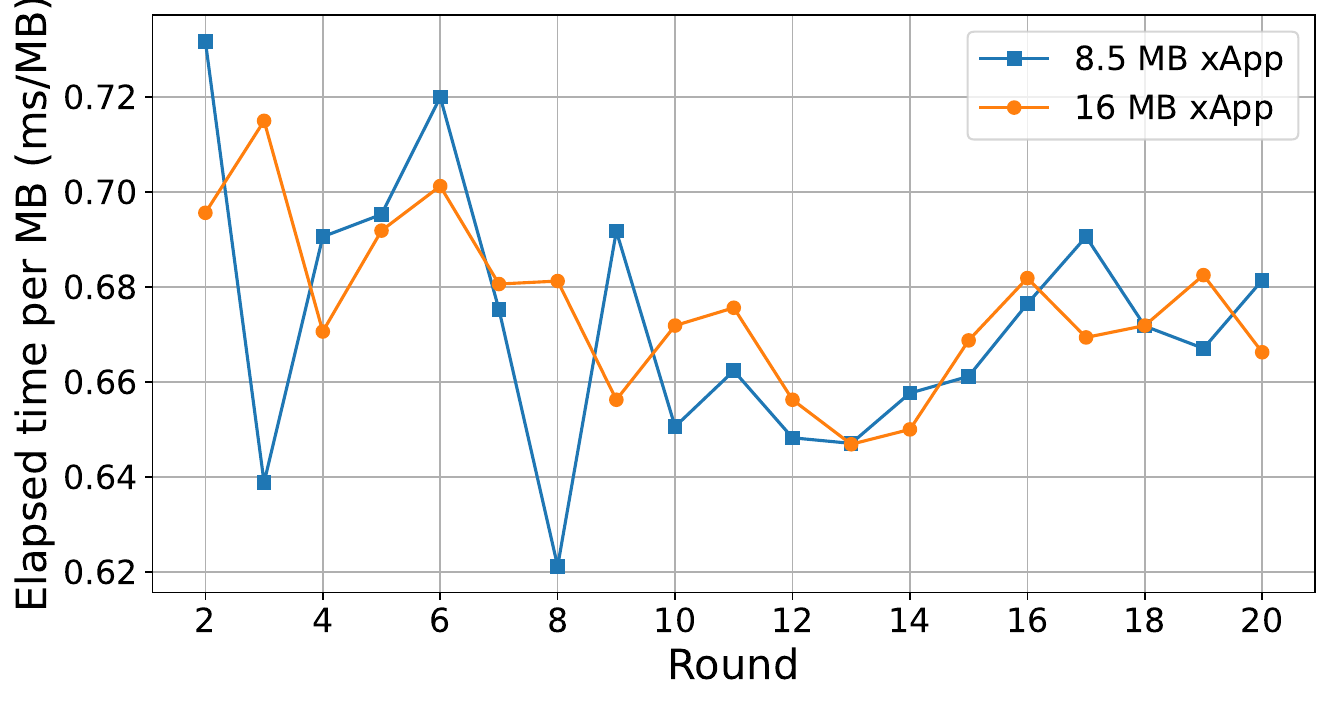}
    \caption{xApp attestation latency per MB of xApp size.}
    \label{fig:attest_2}
\end{figure}

To further analyse this relationship, Figure~\ref{fig:attest_2} presents per-megabyte elapsed time for rounds 2 to 20, excluding the cold-start round. While minor fluctuations occurred in early rounds, per-MB latency converged to approximately 0.67 ms/MByte for both xApp sizes, confirming the linear relationship between input size and hashing time. This linear trend suggests that even substantially larger xApps would remain within feasible attestation time bounds, since the computational cost scales proportionally rather than exponentially. Such scaling behaviour is favourable for deployments where multiple or larger xApps may coexist on COTS hardware.

In summary, the attestation module achieved consistently low, real-time scale latencies, ensuring seamless integration into the near-RT RIC platform without imposing performance bottlenecks. Overall, these results demonstrate that periodic integrity verification is practical in operational Open RAN environments and can be executed at operator-defined intervals without affecting near-RT execution.

\subsection{Test Use Case: Putting It All Together}
\label{testall}
In this section, we evaluate the framework with all three implemented modules integrated. Building on our previous work \cite{10849726}, where we developed the Decision Tree (DT)-based ML model for detecting malicious handover behaviour, we implemented and deployed it as an O-RU-triggered signalling storm detection xApp for this study. This xApp collects KPM reports and uses the trained DT model to detect attacks involving frequent O-RU on/off switching that induces a large number of handovers. The xApp was tested in our Open RAN testbed comprising the RAN emulator and FlexRIC, under the same scenario as the KPM poisoning detection experiments, except we tested here with 50 and 500 UEs..

To evaluate the impact of our safeguarding framework, we tested the signalling storm detection xApp both with and without the integrated safeguards. The xApp was configured to run 100 control loops, each processing KPM records corresponding to a one-second period for the UEs and cells involved in recent handover events, in order to detect malicious handovers and identify malicious cells. In the safeguarded scenario, unlike direct E2SM-KPM subscriptions, the xApp retrieved KPM records from the database, which contained telemetry data verified by the KPM poisoning detection xApp. 

We monitored the timestamps at which the xApp made its decisions to quantify any additional latency introduced by the safeguards. It should be noted that the primary effect of the safeguards is on the time at which the use case xApp can access its required data. Since the control loop operates on a one-second cycle—of which a substantial portion remains idle—the processing delays imposed by the safeguards, which are significantly smaller than one second, do not extend the total execution time of each loop beyond the 1s constraint. Instead, they simply shift the xApp’s operation within its scheduled window, utilising part of its idle time without increasing its aggregated runtime duration over 100 loops. The key requirement, therefore, is not the elimination of delay but ensuring that the imposed latency remains well bounded so that the xApp continues to execute comfortably within its per-cycle timing budget.

Table \ref{tab:usecase} presents the results collected over ten runs, each comprising 100 control loops. As the xApp attestation process operates outside the near-RT control loop and does not impact the operation of the signalling storm detection xApp, it was excluded from this analysis. However, we performed experiments with and without the attestation and did not find any meaningful impact on the performance. For the other two modules, we measured their latencies separately, alongside recording the timestamps of data availability for the xApp under two scenarios: with and without the safeguarding framework implemented.

As shown in the table, the E2 message inspector introduces minimal latency, while the KPM poisoning detection module incurs less than 8 ms and 80 ms in all the scenarios, attributable to its ML-based detection approach. The increase in data-shift time at higher UE density is consistent with the expected growth in telemetry volume and reflects the per-report computational cost of the poisoning detector rather than any structural bottleneck in the framework. Finally, the time shift results indicate that the observed shifts are proportional to the latencies imposed by these modules. However, the minimum and maximum time shifts are not simple sums of the corresponding module latencies, as they occurred across different control loops and runs.

It can be concluded that the proposed defence framework can be integrated into the near-RT RIC environment with latency overheads of less than 8 ms and 80 ms for all the scenarios with 50 and 500 UEs, respectively. This demonstrates the framework’s compatibility with Open RAN near-RT RIC operations while maintaining high efficiency in safeguarding against targeted threats. Overall, the results show that the safeguards operate in a complementary manner, addressing message-, data-, and control-logic–level threats independently while imposing only marginal shifts within the xApp’s execution window, thereby confirming the practicality and scalability of the multi-layer runtime defence design.

\setlength{\tabcolsep}{2pt}
\renewcommand{\tabularxcolumn}[1]{>{\centering\arraybackslash}m{#1}}

\begin{table}[t]
\centering
\caption{Processing times (ms) measured under two UE load scenarios.}
\begin{tabularx}{\columnwidth}{|m{2.8cm}|X|X|X|X|X|X|}
\hline
\multirow{2}{=}{\textbf{Measured Time}} & \multicolumn{3}{c|}{\textbf{50 UEs}} & \multicolumn{3}{c|}{\textbf{500 UEs}} \\ \cline{2-7}
& Min & Max & Avg & Min & Max & Avg \\ \hline
E2 Message Inspector & 0.09 & 0.62 & 0.13 & 0.98 & 5.91 & 1.52 \\ \hline
KPM Poisoning Detection & 6.80 & 7.72 & 7.10 & 71.23 & 78.36 & 74.17 \\ \hline
Data Availability Time Shift & 6.98 & 8.12 & \textbf{7.31} & 73.61 & 79.47 & \textbf{76.74} \\ \hline
\end{tabularx}
\label{tab:usecase}
\end{table}

\section{Discussion and Future Directions}
\label{sec:discussion}

This section critically reflects on the design trade-offs and current limitations of the proposed runtime security components, and outlines directions for future research to enhance their robustness and applicability. While each module addresses specific threat vectors in near-RT RIC environments, further development is needed to extend detection coverage, improve interpretability, and integrate with broader RAN security policies.

It is important to clarify that RF noise, channel fading, mobility dynamics, and other radio-layer variations, including both natural effects and disturbances caused by possible physical-layer attacks, influence only the telemetry used by the data-level poisoning detector. The message-level (E2 inspection) and control-logic-level (xApp attestation) safeguards are not affected by radio conditions because they examine signalling structure and software integrity. This study focuses on runtime security within the near-RT RIC and uses a protocol-level testbed that channel effects, such as radio propagation or physical-layer interference, were not included in it. In practical deployments, radio-layer disturbances, whether natural or adversarial, appear at the near-RT RIC simply as variations in reported KPIs. Handling such variations falls under robustness rather than security from the viewpoint of near-RT RIC operations, and this aspect has been analysed in our previous work on distortion-tolerant learning for Open RAN telemetry \cite{contrastive}, where contrastive augmentation techniques modelled fading-like distortions, Gaussian noise, feature dropout, and other irregularities in KPM reports.

\subsection{Signature-Based E2 Message Inspection}

The E2 message inspection component relies on a signature matching approach, which is well-suited to identifying known malicious signalling patterns embedded in E2 messages. While this method ensures precision and incurs low computational overhead, it is inherently limited in its ability to detect novel or evolving attack strategies that do not match predefined patterns. In particular, an adaptive adversary could deliberately craft protocol-compliant message variants that avoid matching known signatures, a common challenge referred to as \textit{signature evasion}. This limitation is typical of signature-based systems and represents a trade-off between simplicity and robustness. One potential direction for future improvement is the incorporation of learning-based techniques, such as large language models (LLMs) or embedding-based similarity search, to extend detection capabilities to semantically similar or previously unseen message variants, while maintaining compliance with E2AP and E2SM encoding constraints. While such semantic inspection methods could detect a broader class of previously unseen or obfuscated attacks, they would necessarily operate after ASN.1 decoding and therefore incur additional computational overhead and latency. By contrast, the approach presented in this manuscript performs pattern inspection on raw message payloads prior to decoding, which avoids these costs and enables very low-latency operation. The two approaches are thus complementary: semantic methods offer greater expressiveness, whereas pre-decoding inspection provides an efficient first filtering stage.

As discussed in Section~\ref{sec:components}.\ref{sec:E2 message inspection}, the use of naive pattern matching is justified by the short length of typical E2 messages. However, future systems with larger rule sets or more complex injection patterns may benefit from scalable multi-pattern search algorithms such as Aho--Corasick, or from more expressive semantic detectors capable of resisting signature evasion attempts.

\subsection{KPM Poisoning Detection}
The KPM poisoning detection mechanism is based on temporal anomaly detection using an LSTM neural network trained on sequences of historical KPI records. While this approach is effective in identifying subtle deviations and capturing temporal correlations, it inherits limitations common to learning-based detectors. An important one is the dependence on the quality and representativeness of the training data. If the training set does not reflect the full diversity of legitimate operational conditions, the model may suffer from false positives or overlook slow, stealthy poisoning patterns.

Another challenge lies in the interpretability of the LSTM model's decisions. The opaque nature of deep learning-based detection makes it difficult to provide human-understandable justifications for flagged anomalies, which may hinder operator trust and complicate incident response. Exploring explainable AI (XAI) techniques or hybrid approaches that combine statistical thresholds with learned models may help address this limitation.

Lastly, the LSTM model is deployed as a standalone xApp, which simplifies integration but introduces dependencies on telemetry availability, synchronisation, and real-time execution constraints. Additional work is needed to assess its robustness under varying load conditions, data loss, or latency jitter, particularly in resource-constrained RIC environments. Moreover, while the current implementation focuses on detecting a specific type of poisoning attack, the approach can be extended in future work to cover a broader range of poisoning strategies targeting different KPI fields or statistical properties.

\subsection{Runtime xApp Attestation}
While the proposed runtime attestation mechanism enhances xApp integrity verification during execution, several aspects merit further consideration. First, the framework assumes that the attestation logic embedded within the xApp has not itself been compromised. An attacker with control over the xApp may employ code hiding techniques, preserving the original trusted binary image to pass attestation challenges while executing malicious routines elsewhere at runtime. To mitigate this, behavioural analysis techniques—such as monitoring runtime responses or execution patterns—could be explored as a complementary safeguard. These approaches would make it more difficult for attackers to mimic legitimate xApp behaviour, although they introduce additional complexity, require profiling, and may suffer from false positives. Balancing detection accuracy with operational cost remains an open challenge.

Second, the current design does not incorporate hardware-based attestation mechanisms such as Trusted Platform Modules (TPMs) or Trusted Execution Environments (TEEs), which can offer strong integrity assurances. Our intention is not to suggest that such mechanisms are technically difficult to deploy on COTS hardware; rather, their availability cannot be relied upon in Open RAN environments, where near-RT RIC platforms typically run in heterogeneous, virtualised, and cloud-native COTS infrastructures operated by different vendors and cloud providers. This heterogeneity limits the practicality of depending on platform-specific hardware trust anchors as a universal assumption. Lightweight alternatives, such as virtual TPMs, hypervisor-backed attestation, and other software-based trust anchors, represent viable options, but they introduce additional dependencies on platform configuration, virtualisation layers, or cloud-management trust domains. Evaluating these alternatives is an important direction for future work, as our focus in this study is on runtime safeguards that remain deployable without modifications to the underlying hardware or hosting environment.

\subsection{Generalisability to Other RIC Platforms}
\label{sec:general}
While the proposed safeguarding framework has been implemented and evaluated using the FlexRIC platform, its design principles and components are intended to be broadly applicable to other near-RT RIC implementations, such as the O-RAN SC near-RT RIC. The E2 message inspection, KPM poisoning detection, and xApp integrity check modules operate at the levels that are standardised across RIC platforms. However, integrating the framework with other RICs may require adaptation of interfacing components to match platform-specific SDKs, deployment pipelines, and plugin architectures. Future work will include validating the framework on alternative near-RT RIC implementations to further demonstrate its portability and to identify any platform-specific integration challenges.

\subsection{Scalability Considerations}
While the current evaluation includes scenarios with up to 500 UEs, near-RT RIC deployments in dense network environments may involve UE counts exceeding 1000. Based on the observed near-linear scaling behaviour of processing latencies with respect to UE load, we anticipate that the safeguarding framework would continue to operate within acceptable time constraints for higher UE densities. The KPM poisoning detection module, which already utilises batch processing to handle multiple UE records efficiently, is expected to experience proportional latency increases as the number of records per batch grows. To further support scalability, optimisation strategies such as model compression, use of lightweight architectures, or hardware accelerators (e.g. GPU inference) can be explored. Future work will include empirical validation under dense load conditions to confirm these projections and identify any additional architectural enhancements required for large-scale RAN deployments.

\section{Conclusions} 
\label{sec:conclusion}

In this paper, we proposed and evaluated a modular multi-layer reactive defence framework to secure near-RT RIC operations in Open RAN environments. By categorising runtime threats into message-level, data-level, and control logic-level dimensions, we designed and implemented targeted safeguarding modules comprising a signature-based E2 message inspector, an LSTM-based KPM poisoning detector, and a runtime hashing-based xApp attestation mechanism. Our experimental evaluation on an Open RAN testbed demonstrated that the proposed framework can achieve high detection accuracy with minimal latency overheads, remaining well within near-RT time constraints even under increased UE loads. This confirms its suitability for integration into practical Open RAN deployments. Furthermore, the generalisability of our approach, both across different near-RT RIC platforms and as a structured guideline for categorising threats and placing safeguards, provides a foundation for future extensions. Future work will extend detection capabilities to cover a broader range of poisoning attacks, explore learning-based E2 message inspection techniques, and behavioural attestation. Overall, this study provides a step towards layered, runtime-aware defence mechanisms to enhance the resilience and trustworthiness of Open RAN systems.

 \section*{Acknowledgment}
This work has been supported in part by the ORAN-TWIN-X subproject under CHEDDAR: Communications Hub for Empowering Distributed Cloud Computing Applications and Research funded by the UK Engineering and Physical Sciences Research Council (EPSRC) under grant numbers EP/Y037421/1 and EP/X040518/1 and in part by Abu Dhabi University’s Office of Sponsored Programs in the United Arab Emirates (Grant number: 19300917).

\bibliographystyle{IEEEtran}
\bibliography{references}

\begin{IEEEbiography}
[{\includegraphics[width=1in,height=1.25in,clip,keepaspectratio]{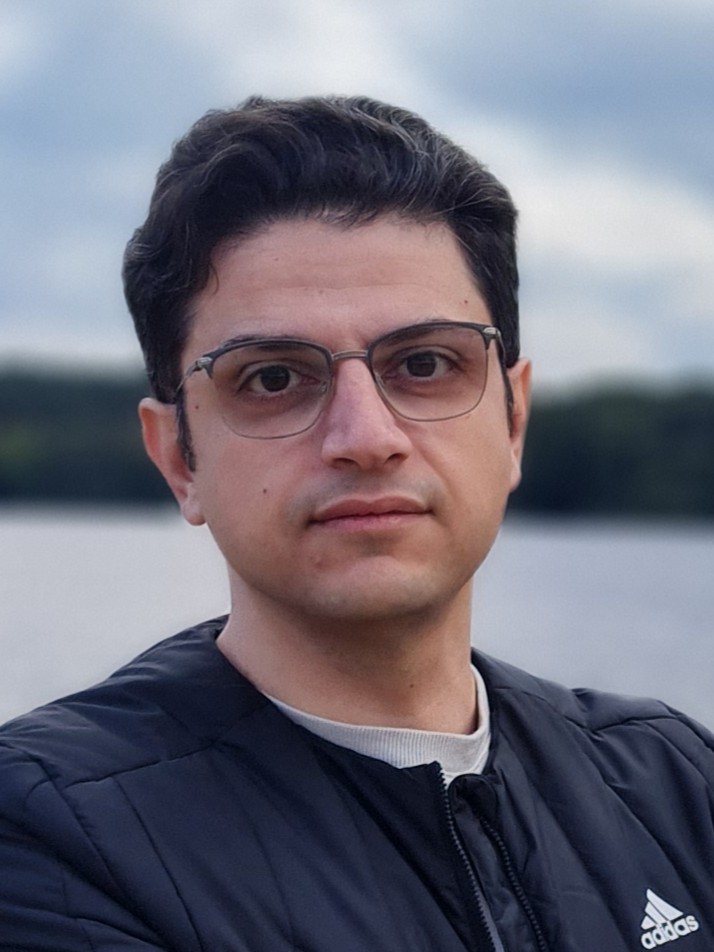}}]{Hamed Alimohammadi} ~is a Senior Research Fellow in Mobile Networks at the 6G Innovation Centre (6GIC), Institute for Communication Systems (ICS), University of Surrey, specialising in Open RAN self-organisation and security. He previously spent nearly a decade in mobile network governance and engineering roles. He received his BSc and MSc degrees in Computer Engineering from the Ardabil and Science and Research (Tehran) branches of Azad University, Iran, respectively. He obtained his PhD in Computer Engineering from Razi University, Kermanshah, Iran, in 2020, with a focus on software-defined networking. At Surrey, he has contributed to flagship research programmes including 6G-SMART and HiPer-RAN, and was part of the team recognised with the UK Government’s \textit{Future Network Incremental Innovative Award} in 2025. His research interests include Open RAN security, machine learning for self-organising networks, and high-performance computing.
\end{IEEEbiography}

\begin{IEEEbiography}
[{\includegraphics[width=1in,height=1.25in,clip,keepaspectratio]{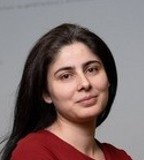}}]{Samara Mayhoub} ~is a Research Fellow with the Centre for Secure Information Technologies (CSIT), Queen’s University Belfast, U.K. She was a Postdoctoral Researcher with the 6G Innovation Centre (6GIC), University of Surrey, U.K., in 2024, where she conducted research on cybersecurity for O-RAN networks. She received the Ph.D. in Network Security from Samara National Research University, Russia, in 2022, and the M.Sc. in Information Systems and Technologies from MIREA—Russian Technological University, Moscow, Russia, in 2018. She also served as a Lecturer at the Department of supercomputers and computer science, Samara National Research University, in 2023.
\end{IEEEbiography}

\begin{IEEEbiography}
[{\includegraphics[width=1in,height=1.25in,clip,keepaspectratio]{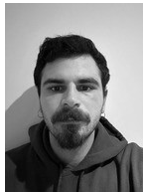}}]{Sotiris Chatzimiltis} ~(Graduate Student Member, IEEE) received the B.Sc. degree in computer science from the University of Cyprus, Cyprus, in 2021, and the M.Sc. degree in computer vision, machine learning and robotics from the University of Surrey, Guildford, U.K., in 2022. He is currently working toward the Ph.D. degree with the Institute for Communication Systems (ICS), the University of Surrey. His current research interests include machine learning, intrusion detection systems, distributed AI and Open RAN security.
\end{IEEEbiography}

\begin{IEEEbiography}
[{\includegraphics[width=1in,height=1.25in,clip,keepaspectratio]{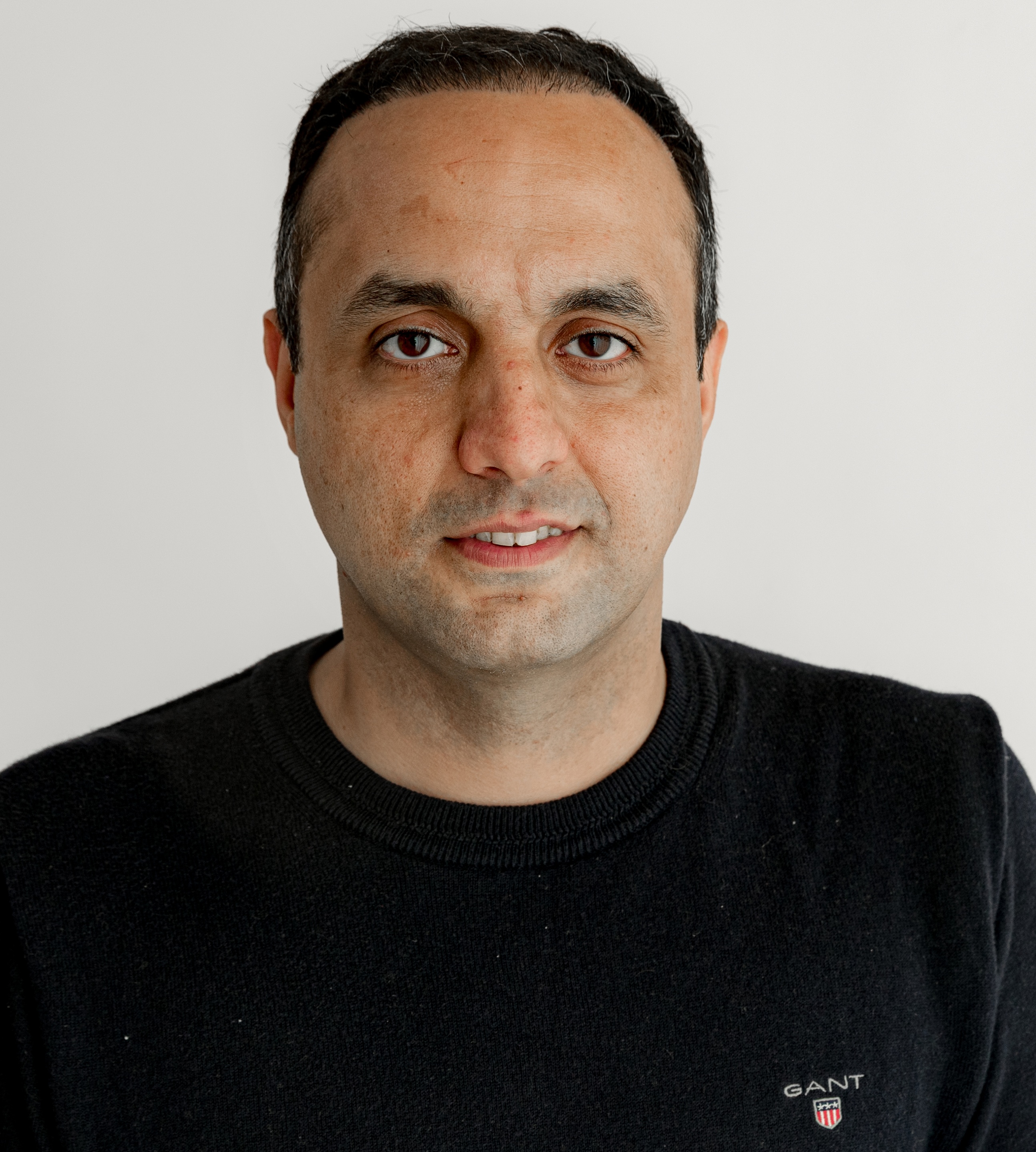}}]{Mohammad Shojafar} \textbf{ (M'17-SM'19)} is an Associate Professor in network security and an Intel Innovator, professional ACM member and ACM distinguished speaker, a senior member of IEEE, a fellow of the Higher Education Academy (FHEA), and a Marie Curie alumnus, working in the 6G Innovation Centre (6GIC), Institute for Communication Systems (ICS), at the University of Surrey, UK. Before joining 6GIC, he was a senior researcher and a Marie Curie fellow in the SPRITZ Security and Privacy Research group at the University of Padua, Italy. Dr Mohammad secured around $\pounds$2M as PI in various EU/UK projects. He is an associate editor in \textit{IEEE Transactions on Network and Service Management}, \textit{IEEE Transactions on Intelligent Transportation Systems}, \textit{IEEE Transactions on Green Communications and Networking}, \textit{IEEE Transactions on Consumer Electronics}, and \textit{Computer Networks}. He served as the Open RAN Workshop Chair at IEEE WCNC 2025 and IEEE EUCNC 2025, as Publicity Chair for IEEE WiSec 2023, and as a Co-chair for IEEE TrustCom 2021. In 2023, he chaired the IEEE Communications Society-supported 6G Security and Privacy Workshop named “6GICCLICK (6GSEC)”. He is also an active member of the ETSI Intelligent Transportation Systems Group, ETSI Network Functions Virtualisation Group, GSMA Open-Telco LLM Group, and the 3GPP 5G Service and System Aspects (SA) working group. He has contributed to standardisation activities of the O-RAN Alliance as part of Work Group 11 in 2024 and has been serving on the IEEE Senior Membership Review Panel since 2024. Dr. Shojafar and the Surrey team received the “Incremental Innovative” Future Network Awards for the DSIT/UKTIN HiPer-RAN Project from the UK Government in 2025 and the Best Paper Award at the 7th IEEE Cyber Security in Networking Conference (CSNet) in 2023. He is also the author of three recently published books on cybersecurity and network security with Springer.
\end{IEEEbiography}

\begin{IEEEbiography}
[{\includegraphics[width=1in,height=1.25in,clip,keepaspectratio]{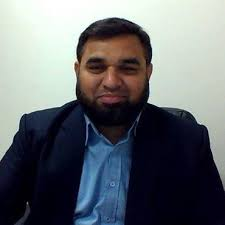}}]{Muhammad Nasir Mumtaz Bhutta} ~is an Assistant Professor of Computer Science and Information Technology at Abu Dhabi University, specialising in network security, blockchain technology, and the Internet of Things (IoT). With a Ph.D. from the University of Surrey, his research focuses on developing secure, privacy-preserving frameworks using machine learning, federated learning, and deep learning to protect decentralised systems and smart city infrastructures. His work is supported by significant research funding from the Deanship of Scientific Research at King Faisal University and the Abu Dhabi University Research Office, specifically advancing areas such as big data analytics, agricultural supply chain tracking, and secure multi-party computation. Additionally, he has contributed to high-impact international projects funded by the European Space Agency and the EPSRC UK, aiming to democratise secure and privacy-aware artificial intelligence.
\end{IEEEbiography}

\end{document}